\documentclass[fleqn,10pt]{wlscirep}
\usepackage{color}
\usepackage[utf8]{inputenc}
\usepackage[T1]{fontenc}
\usepackage{graphicx}%
\usepackage{multirow}%
\usepackage{amsmath,amssymb,amsfonts}%
\usepackage{amsthm}%
\usepackage{mathrsfs}%
\usepackage[title]{appendix}%
\usepackage{xcolor}%
\usepackage{textcomp}%
\usepackage{manyfoot}%
\usepackage{booktabs}%
\usepackage{algorithm}%
\usepackage{algorithmicx}%
\usepackage{algpseudocode}%
\usepackage{listings}%
\usepackage{caption}
\usepackage{bm}  
\usepackage{float}
\usepackage{hyperref}

\usepackage{caption} 
\usepackage{hyperref} 

\newcounter{edf} 

\newenvironment{edfigure}{
    \refstepcounter{edf}
    \begingroup
    \renewcommand{\figurename}{Extended Data Fig.}%
    \begin{figure}[htbp]
}{
    \end{figure}
    \endgroup
}

\title{Simultaneous super-resolution and optical sectioning with four-beam interference structured illumination microscopy (4I-SIM)}

\author[1,2,3,4,${\dag}$]{Jiaming Qian}
\author[1,2,3,${\dag}$]{Jing Feng}
\author[1,2,3,4,${\dag}$]{Hongjun Wu}
\author[1,2,3]{Maoxian Zhang}
\author[1,2,3]{Dongqin Lu}
\author[1,2,3]{Tianchi Kang}
\author[1,2,3]{Xinyu Han}
\author[3,4,**]{Qian Chen}
\author[1,2,3,4*]{Chao Zuo}

\affil[1]{Smart Computational Imaging (SCI) Laboratory, Nanjing University of Science and Technology, Nanjing, Jiangsu Province 210094, China}
\affil[2]{Smart Computational Imaging Research Institute (SCIRI) of Nanjing University of Science and Technology, Nanjing, Jiangsu Province 210094, China}
\affil[3]{Jiangsu Key Laboratory of Visual Sensing $\&$ Intelligent Perception, Nanjing, Jiangsu Province 210094, China}
\affil[4]{State Key Laboratory of Extreme Environment Optoelectronic Dynamic Measurement Technology and Instrument, Taiyuan, Shanxi Province 030051, China}

\affil[*]{zuochao@njust.edu.cn}
\affil[**]{chenqian@njust.edu.cn}
\affil[$\dag$]{These authors contributed equally to this work}

\begin{abstract}
Structured illumination microscopy (SIM) has emerged as a widely adopted super-resolution fluorescence imaging modality, offering high speed, low phototoxicity, large field-of-view, and compatibility with conventional probes. However, when applied to thick or scattering specimens, conventional two-dimensional SIM (2D-SIM) suffers from the ``missing cone'' problem in its optical transfer function, resulting in prominent out-of-focus background and severe reconstruction artifacts that compromise image fidelity. Here, we present four-beam interference structured illumination microscopy (4I-SIM), which introduces additional interference orders to expand lateral frequency support and compensate the axial missing cone simultaneously. This strategy achieves artifact-free super-resolution with intrinsic optical sectioning, effectively overcoming the fundamental limitation of 2D-SIM without additional acquisition overhead. Experimental validation across diverse thick fixed and live specimens demonstrates that 4I-SIM delivers nearly twofold lateral resolution enhancement and substantially improved sectioning compared with its 2D counterpart, achieving lateral and axial resolutions of 103 nm and 336 nm, respectively. In particular, 4I-SIM reveals mitochondrial remodeling and apoptosis under high-glucose stress with millisecond temporal resolution -- features that remain obscured with conventional SIM. With minimal hardware modification, low phototoxicity, and open-source reconstruction tools, 4I-SIM establishes a practical and reproducible platform for simultaneous super-resolution and optical sectioning imaging in complex biological environments.
\end{abstract}

\begin{document}
\renewcommand{\figurename}{Fig.}
\flushbottom
\maketitle
%
%
\thispagestyle{empty}
\section{Introduction}\label{sec1}

Understanding the nanoscale architecture and rapid dynamics of subcellular organelles, such as mitochondria, endoplasmic reticulum, and cytoskeletal filaments, is essential for elucidating fundamental biological processes and disease mechanisms \cite{newmeyer2003mitochondria,akhmanova2008tracking,bosch2014structural,behnke1964electron}. Many of these processes, including mitochondrial fission and fusion, endocytic trafficking, and cytoskeletal remodeling, occur at spatiotemporal scales well below the diffraction limit and often unfold on sub-second to millisecond timescales. Capturing such fast, nanoscale events in their native physiological context therefore demands live-cell imaging modalities that provide high spatial resolution, millisecond-level temporal fidelity, and minimal phototoxicity \cite{westphal2008video, chen2007mitochondrial,cogliati2013mitochondrial}. Structured illumination microscopy (SIM), which offers up to twofold lateral resolution enhancement while preserving photon efficiency and compatibility with standard fluorescent probes, has consequently become a widely adopted modality for long-term, high-speed visualization of subcellular dynamics in living cells \cite{gustafsson2000surpassing,heintzmann1999laterally,li2015extended,guo2018visualizing,huang2018fast,zhao2022sparse,wen2021high,wang2022high,zhang2023deep,chen2024self}.

Despite its broad utility, conventional two-dimensional (2D) super-resolution structured illumination microscopy (SR-SIM) is often compromised in thick or scattering biological specimens, where out-of-focus fluorescence degrades image contrast and obscures fine structural details. This limitation stems from the well-known axial missing cone in its three-dimensional (3D) optical transfer function (OTF), which renders essential axial spatial frequencies inaccessible and allows out-of-focus fluorescence to dominate the detected signal \cite{macias1988missing,wicker2013phase}. The resulting loss of axial support produces blurred backgrounds, distorted structures, and reconstruction artifacts that severely degrade sectioning capability and undermine image fidelity in complex or densely labeled environments. These effects are particularly consequential for live-cell imaging, where axial background can obscure dynamic subcellular processes and undermine the reliability of structural interpretation. 3D-SIM addresses this limitation by employing coherent three-beam interference to generate volumetric illumination patterns that extend axial frequency support, enabling simultaneous lateral and axial resolution enhancement \cite{gustafsson2008three,schermelleh2008subdiffraction,cao2023open,cao2025fast}. Nevertheless, these advantages come at the cost of extensive z-stack acquisition, stringent optical alignment, increased illumination dose, and reduced temporal resolution -- trade-offs that restrict its practicality for general-purpose biological imaging, especially in dynamic or phototoxicity-sensitive live-cell contexts.

As a simplified alternative, optical-sectioning SIM (OS-SIM) suppresses out-of-focus fluorescence by exploiting the intrinsic loss of modulation contrast with defocus under spatially incoherent illumination \cite{neil1997method,lim2008wide,santos2009optically,mertz2010scanning,patorski2014optically,karadaglic2008image,karadaglic2008image2}. Building on this concept, multiple strategies, including WLR-SIM \cite{o2014optimized} and iSIM \cite{dan2020super}, have been developed to integrate OS into SR-SIM. However, these approaches typically exploit illumination frequencies near 1.5$\times$ the original cutoff to partially mitigate the axial missing cone, which inherently restricts the heterodyning range and leads to a substantial loss of lateral SR compared with standard SR-SIM. 
Deconvolution-based methods applied to volumetric stacks, such as BF-SIM \cite{mo2023quantitative}, attenuate defocus contributions but inherit the heavy acquisition load and phototoxicity of 3D-SIM. Other efforts, inspired by HiLo microscopy \cite{lim2008wide}, attempt to enhance OS by selectively suppressing out-of-focus spectral components and reinforcing in-focus modulated signals \cite{johnson2020artifact, wen2021high}. More recently, model-free post-processing techniques such as Dark-SIM \cite{cao2024dark} have emerged, leveraging statistical priors to remove background without requiring additional hardware. However, operating solely at the post-processing stage, these methods do not modify the underlying encoding of 3D information during image formation and therefore cannot recover axial spatial frequencies fundamentally lost to the missing cone, intrinsically limiting OS and compromising the fidelity of weak but genuine structures in dense or thick specimens. Collectively, these limitations highlight a persistent methodological gap: there is still no SIM modality capable of delivering both robust OS and $ \sim $two-fold lateral SR while preserving the simplicity, speed, and live-cell compatibility of conventional 2D-SIM -- a capability essential for high-fidelity imaging in thick or heterogeneous specimens.

Here, we present four-beam interference structured illumination microscopy (4I-SIM), a composite illumination and reconstruction framework that bridges the long-standing gap between robust OS and high lateral SR within a standard 2D SIM architecture. By reshaping the effective 3D OTF through engineered four-beam interference, 4I-SIM generates additional illumination orders that substantially extend the accessible lateral spatial-frequency support while simultaneously compensating for the axial missing cone without resorting to volumetric illumination patterns or z-stack acquisition. To fully exploit the enriched frequency content, we develop a 3D OTF-informed optimal frequency-weighting reconstruction framework that suppresses out-of-focus contributions while reinforcing weak axial-frequency components, enabling intrinsic OS alongside high lateral fidelity even under conditions where conventional 2D SIM fails. Importantly, 4I-SIM preserves the standard nine-frame acquisition and requires no additional hardware or alignment, ensuring seamless compatibility with existing SIM platforms. Across a variety of thick fixed and live specimens, 4I-SIM consistently delivers effective OS while retaining the lateral resolution expected from standard SR-SIM. In BSC-1 cells imaged with a 100$\times$/1.45 NA objective, 4I-SIM reached lateral and axial resolutions of 103.4 nm and 336.2 nm, representing $ \sim $1.91$\times$ and $ \sim $1.59$\times$ improvements over wide-field imaging. In heterogeneous live-cell environments, such as high-glucose-stressed cultures, 4I-SIM further revealed mitochondrial remodeling and apoptosis dynamics that remain obscured with conventional SI due to its lack of sectioning capability. These results establish 4I-SIM as a practical, robust, and broadly applicable solution for high-fidelity SR imaging across diverse and complex biological contexts.

\section{Results}\label{sec2}

\subsection{3D OTF reinterpretation of 2D SIM}

Conventional 2D-SIM is most often formulated as a lateral SR technique, in which high-frequency information is recovered through heterodyne mixing between the object spectrum and structured illumination patterns produced by two-beam interference (Figs. \ref{fig:principle}a1). As a result, in the frequency domain, the recorded signal is decomposed into a set of modulated spectral components, as illustrated in Fig. \ref{fig:principle}b1. These components comprise the central 0-order term \( C_{0} \), together with the $\pm$1st-order terms \( C_{\pm 1_{\mathrm{Conv.}}} \) symmetrically distributed on either side, where \( C_{\pm 1_{\mathrm{Conv.}}} \) carries SR information beyond the diffraction limit by approximately the illumination frequency $\mathbf{k}_{\text{ex}}$ (the subscript ${Conv.}$ is used to distinguish the conventional illumination mode). In this framework, axial effects are typically treated implicitly, and OS is regarded as a secondary consequence of modulation contrast rather than a property of the underlying transfer function.

From a physical imaging standpoint, however, SIM is inherently a 3D process. When imaging a thick specimen, the interaction between structured illumination and the object encodes volumetric information that is transferred to the recorded images through the 3D OTF. As a result, the effective 2D transfer function governing 2D-SIM imaging corresponds to a projection of the 3D OTF along the axial frequency dimension \cite{1986On, 1991Three}, given by:
\begin{equation}
	\tilde{h}(\mathbf{k}_{xy}) = \int \tilde{h}(\mathbf{k})\, \mathrm{d}{\mathbf{k}_z}
	\label{eq2}
\end{equation}
where $\tilde{h}(\mathbf{k})$ denotes the 3D OTF, defined over spatial frequency coordinates (\(\mathbf{k}_{x}\), \(\mathbf{k}_{y}\), \(\mathbf{k}_{z}\)). 
This projection redistributes the available axial-frequency content into the lateral frequency plane, disproportionately accumulating low-axial-frequency components while suppressing high-axial-frequency contributions.
Viewed from this perspective, the limited OS of conventional 2D-SIM arises from the anisotropic support of its effective 3D OTF. In particular, the axial missing cone leads to a systematic under-representation of axial spatial frequencies, causing out-of-focus and scattered contributions to be preferentially transferred into the reconstructed image (Supplementary Figs. S1a and S1b). Importantly, this limitation is intrinsic to how spatial frequencies are encoded and weighted during image formation, rather than a consequence of reconstruction algorithms or post-processing choices.
This 3D OTF reinterpretation reveals a critical but previously underexploited degree of freedom in 2D-SIM. Although standard SIM acquisitions are not designed for volumetric imaging, the frequency-mixing process inherently generates multiple spectral components with distinct axial sensitivities. In conventional reconstructions, these components are effectively collapsed into a single projected transfer function, discarding their differential axial information  (detailed in Supplementary Notes S1 and S2).

Recognizing OS as an encoding- and transfer-function-level problem fundamentally changes how it can be addressed in 2D-SIM. Rather than relying on background suppression or volumetric illumination, improving sectioning performance requires reshaping the effective 3D OTF by selectively accessing and combining frequency components according to their axial information content. This insight provides the conceptual foundation for four-beam interference structured illumination microscopy (4I-SIM), which explicitly treats OS as a problem of 3D transfer-function engineering within a standard 2D-SIM acquisition framework.

\subsection{Four-beam interference and asynchronous phase encoding for OTF engineering}

\begin{figure}[htb]
	\centering
	\includegraphics[width=1\textwidth]{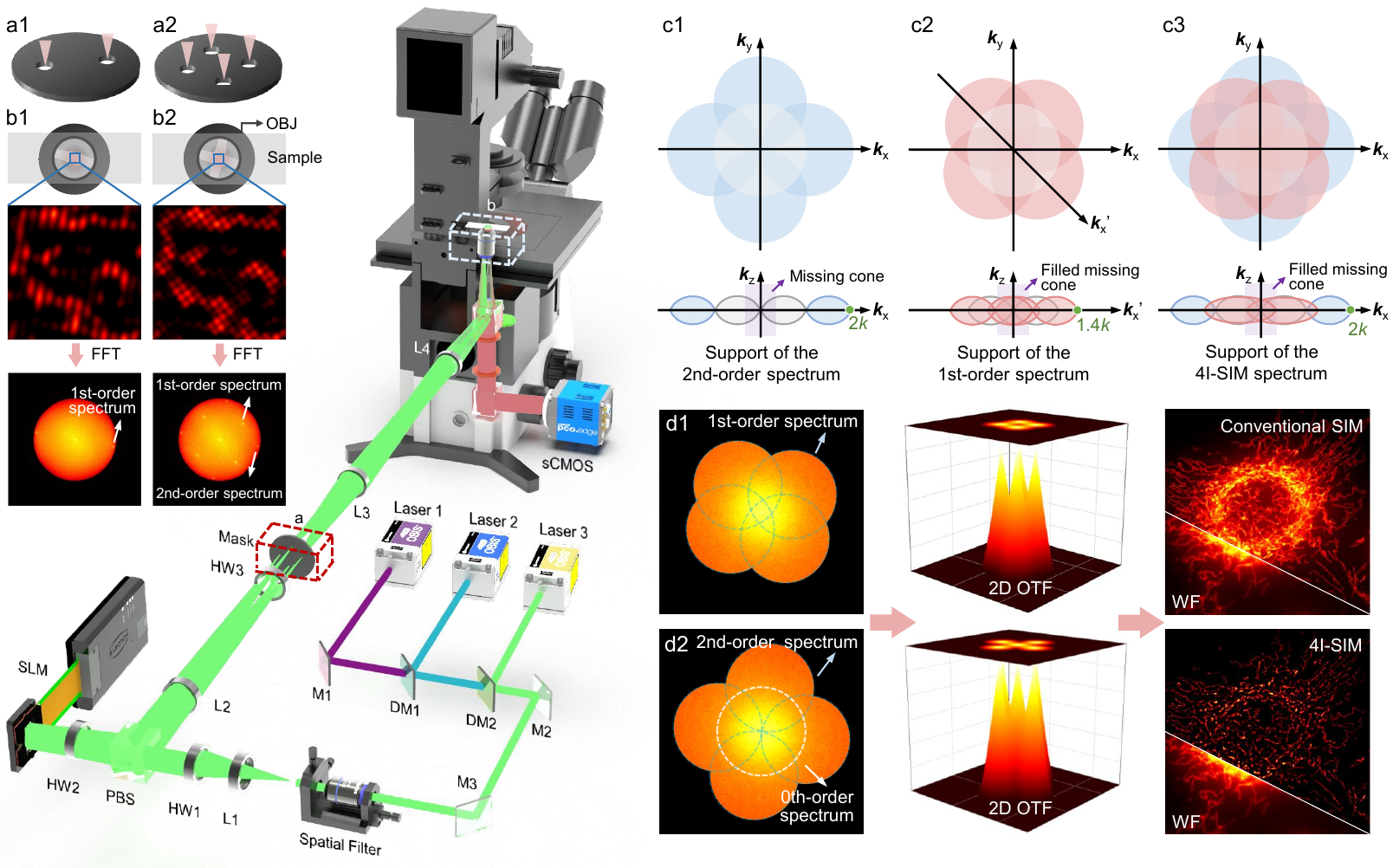}
	\caption{\textbf{Schematic illustration of the 4I-SIM principle.} 
		\textbf{a1} Two-beam interference configuration in conventional 2D SR-SIM.  
		\textbf{a2} Four-beam interference configuration in 4I-SIM. 
		\textbf{b1} The initial illumination image and its spectrum map in conventional 2D SR-SIM. 
		\textbf{b2} The initial illumination image and its spectrum map  in 4I-SIM. 
		\textbf{c} The support regions of different modulated spectral components in 4I-SIM, including the second-order spectral support region which suffers from the missing-cone effect (\textbf{c1}), the first-order spectral support region where the missing cone is filled (\textbf{c2}), and the combined support region of both first- and second-order components (\textbf{c3}).
        \textbf{d} Workflow of the frequency-domain composite filtering strategy constrained by the 3D OTF, where \textbf{d1} shows the effective 2D OTF applied to the first-order spectral components, used to construct a Wiener filter that enhances the axial response, and \textbf{d2} shows the effective 2D OTF applied to the second-order spectral components, used to construct a Wiener filter that suppresses low-frequency components in the missing-cone region.
	}
	\label{fig:principle}
\end{figure}

Guided by the 3D OTF reinterpretation of SIM, 4I-SIM employs coherent four-beam interference to actively engineer the frequency content of the illumination field (Fig. \ref{fig:principle}a2). In contrast to conventional two-beam SIM, four-beam interference generates a composite structured illumination pattern containing multiple spatial-frequency components with distinct axial transfer characteristics (Fig. \ref{fig:principle}b2). The resulting illumination spectrum $\tilde D $ can be expressed in the frequency domain as a superposition of zero-, first-, and second-order components:
\begin{equation}
\tilde D({\bf{k}}) = \;O({\bf{k}}){\mkern 1mu} {{\tilde S}_0}({\bf{k}}) + \sum\nolimits_{i =  \pm 1} {\frac{{m_{u,v}^1}}{2}O({\bf{k}}){\mkern 1mu} {{\tilde S}_i}\left[ {{\bf{k}} - i \cdot ({{\bf{k}}_u} \pm {{\bf{k}}_v})} \right]}  + \sum\nolimits_{i =  \pm 2} {\frac{{m_{u,v}^2}}{2}O({\bf{k}}){\mkern 1mu} {{\tilde S}_i}\left[ {{\bf{k}} - i \cdot {{\bf{k}}_{u,v}})} \right]} 
\end{equation}
where $\mathbf{k}$ denotes the spatial-frequency vector, $O(\mathbf{k})$ and $\tilde{S}(\mathbf{k})$ represent the object spectrum and the system OTF, respectively, the index $i$ represents the $i$-th spectral order, $\mathbf{k}_u$ and $\mathbf{k}_v$ are the spatial wave vectors of the horizontal and vertical illumination patterns, and $m_u^i$, $m_v^i$ denote the modulation depth for the $i$-th spectral order in the horizontal and axial directions, respectively. Among these spectral distributions, second-order components function similarly to \( C_{\pm 1_{Conv.}} \) in conventional illumination, primarily extending lateral frequency support nearly twofold, as shown in Fig. \ref{fig:principle}c1. Although the first-order components provide only $\sim1.4\times$ lateral resolution enhancement, they partially fill in the diagonal spectral gaps left by the synthetic second-order components, improving the uniformity of lateral resolution across the entire field (Figs. \ref{fig:principle}c2 and \ref{fig:principle}c3). Importantly,  the first-order components exhibit enhanced axial sensitivity from the perspective of the 3D OTF (Figs. \ref{fig:principle}c2 and \ref{fig:principle}c3). As a result, the four-beam interference pattern simultaneously provides frequency components that favor lateral SR and components that compensate for the axial missing cone. This complementary axial-lateral frequency distribution forms the basis for intrinsic OS within a standard 2D SIM acquisition. More technical details are provided in Supplementary Note S3, Supplementary Figs. S1c-S1e and Supplementary Video S1.

To reliably separate the multiple frequency components generated by four-beam interference, 4I-SIM adopts an asynchronous phase-encoding strategy  (detailed in Supplementary Table S1). Unlike conventional SIM, which employs synchronized phase stepping across all illumination components, asynchronous phase shifting decouples the phase evolution of different beam pairs, increasing the diversity of frequency encoding without increasing the number of raw images. The phase-modulated raw images can be written in matrix form as:
\begin{equation}
{{\tilde D}_n}({\bf{k}}) = {\left[ {\begin{array}{*{20}{c}}
1\\
{m_u^1  {e^{ - j({\varphi _u} - {\varphi _v} + \varphi _u^1)}}/2}\\
{m_u^1  {e^{j({\varphi _u} - {\varphi _v} + \varphi _u^1)}}/2}\\
{m_v^1  {e^{ - j({\varphi _u} + {\varphi _v} - \pi  + \varphi _v^1)}}/2}\\
{m_v^1  {e^{j({\varphi _u} + {\varphi _n} - \pi  + \varphi _v^1)}}/2}\\
{m_u^2  {e^{ - j(2{\varphi _u} - \pi  + \varphi _u^2)}}/2}\\
{m_u^2  {e^{j(2{\varphi _u} - \pi  + \varphi _u^2)}}/2}\\
{m_v^2  {e^{ - j(2{\varphi _v} - \pi  + \varphi _v^2)}}/2}\\
{m_v^2  {e^{j(2{\varphi _v} - \pi  + \varphi _v^2)}}/2}
\end{array}} \right]^T}\left[ {\begin{array}{*{20}{c}}
{O({\bf{k}}){\mkern 1mu} {{\tilde S}_0}({\bf{k}})}\\
{O({\bf{k}}){\mkern 1mu} {{\tilde S}_{ - 1}}[{\bf{k}} + ({{\bf{k}}_u} - {{\bf{k}}_v})]}\\
{O({\bf{k}}){\mkern 1mu} {{\tilde S}_{ + 1}}[{\bf{k}} - ({{\bf{k}}_u} - {{\bf{k}}_v})]}\\
{O({\bf{k}}){\mkern 1mu} {{\tilde S}_{ - 1}}[{\bf{k}} + ({{\bf{k}}_u} + {{\bf{k}}_v})]}\\
{O({\bf{k}}){\mkern 1mu} {{\tilde S}_{ + 1}}[{\bf{k}} - ({{\bf{k}}_u} + {{\bf{k}}_v})]}\\
{O({\bf{k}}){\mkern 1mu} {{\tilde S}_{ - 2}}({\bf{k}} + 2{{\bf{k}}_u})}\\
{O({\bf{k}}){\mkern 1mu} {{\tilde S}_{ + 2}}({\bf{k}} - 2{{\bf{k}}_u})}\\
{O({\bf{k}}){\mkern 1mu} {{\tilde S}_{ - 2}}({\bf{k}} + 2{{\bf{k}}_v})}\\
{O({\bf{k}}){\mkern 1mu} {{\tilde S}_{ + 2}}({\bf{k}} - 2{{\bf{k}}_v})}
\end{array}} \right]
\end{equation}
where \( T \) denotes the transpose operation, \( n \) represents the image sequence index, where \( n \in \{1, 2, \dots, 9\} \), \( \varphi_u \) and \( \varphi_v \) are the phase shifts of the horizontal and vertical illumination patterns, respectively (values taken from Supplementary Table S1), and \( \varphi_{u,v}^{1,2} \) denote the initial phases corresponding to different modulation orders. By combining the set phase shifts $\varphi_u$ and $\varphi_v$, the illumination parameter estimation method based on principal component analysis (PCA) \cite{qian2023structured} allows for the accurate retrieval of experimental parameters such as wave vectors, initial phases, and modulation depths, thus enabling precise spectral separation through matrix inversion (detailed in Supplementary Notes S3 and S4). This asynchronous encoding strategy enables robust spectral separation using a standard nine-frame acquisition, equivalent to conventional 2D SR-SIM. By permitting larger phase increments and relaxing phase synchronization constraints, the approach improves numerical stability and reduces sensitivity to phase errors, while preserving compatibility with existing SIM hardware.

From the perspective of OTF engineering, asynchronous phase encoding ensures that the enriched frequency components generated by four-beam interference are faithfully separated and preserved for subsequent reconstruction. Together, four-beam interference and asynchronous phase encoding establish a flexible and efficient mechanism for reshaping the effective 3D OTF without resorting to volumetric illumination or axial scanning.

\begin{figure}[htb]
	\centering
	\includegraphics[width=1\textwidth]{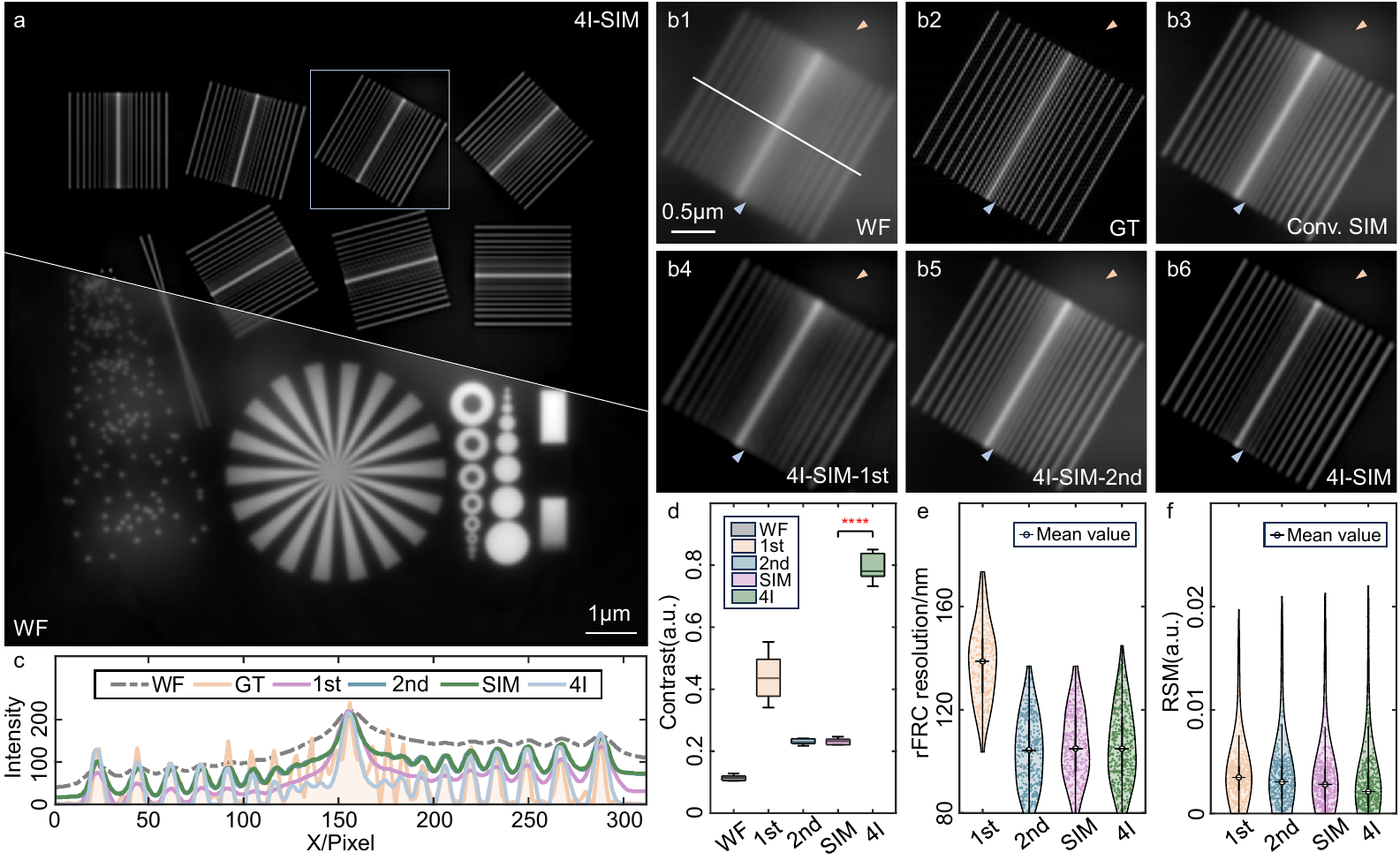}
	\caption{\textbf{Comparative simulations on standard structural samples with out-of-focus signals.} 
		\textbf{a} Comparison of the wide-field image and the super-resolution image obtained by 4I-SIM. 
		\textbf{b} Magnified wide-field image and super-resolution images from the blue boxed regions in \textbf{a} obtained by different methods (first-order spectrum reconstruction of 4I-SIM, second-order spectrum reconstruction of 4I-SIM, conventional SR-SIM, 4I-SIM and the ground truth).
		\textbf{c} Fluorescence intensity profiles along the white line in \textbf{b}. 
		\textbf{d} Comparison of local image contrast between the wide-field image and super-resolution images obtained by different methods.
		\textbf{e} Comparison of rFRC lateral resolution distributions between super-resolution images obtained by different methods. 
		\textbf{f} Comparison of RSM distributions between super-resolution images obtained by different methods. `WF' represents `wide-field', `GT' means `ground truth', `1st' and `2nd' are the first- and second-order spectrum reconstructions of 4I-SIM respectively, `Conv. SIM' / `SIM' here specifically refers to the conventional SR-SIM, and `4I' means `4I-SIM'.A two-tailed paired Student's $t$-test was performed on the contrast values in \textbf{d}, with **** indicating $p<0.0001$. 
		Each simulation was independently repeated ten times with consistent results. Colored arrows highlight regions with significant reconstruction differences. 
		Scale bars: 1 $\mu$m (\textbf{a}); 500 nm (\textbf{b}).}
	\label{fig:simulation}
\end{figure}

\subsection{3D OTF-informed optimal frequency-weighting reconstruction framework}

Following spectral separation, the central challenge of 4I-SIM reconstruction lies in how the extracted frequency components are optimally combined to achieve both high lateral fidelity and robust OS. In conventional SR-SIM, spectral recombination is commonly performed using generalized Wiener filtering, in which different frequency components are weighted according to their corresponding 2D OTF responses. While effective for lateral SR, this strategy becomes suboptimal when applied to 4I-SIM. From the 3D OTF perspective, the limitation of Wiener-based fusion arises from the projection of volumetric frequency content onto the lateral frequency plane. As discussed above, the effective 2D OTF is obtained by integrating the 3D OTF along the axial frequency dimension (Eq. \ref{eq2}), causing frequency components originating from the axial missing-cone region to be disproportionately mapped onto the central region of the 2D frequency plane. Consequently, conventional Wiener filtering assigns relatively large weights to frequency components affected by missing-cone deficiencies, while underweighting components that intrinsically compensate for axial-frequency loss. This mismatch leads to ineffective OS and residual out-of-focus background.

In contrast, 4I-SIM treats reconstruction as an optimal frequency-selection problem informed by the 3D OTF. Rather than assigning weights solely based on lateral frequency magnitude, the reconstruction explicitly accounts for the axial transfer characteristics of each spectral component. As illustrated in Extended Data Figs. \ref{EDF0}a1 and \ref{EDF0}b1, both the zeroth- and second-order spectral components suffer from missing-cone deficiencies in the 3D OTF, and their corresponding projections onto the 2D Fourier plane should, in principle, be suppressed. However, these projected components coincide with regions where the effective 2D OTF exhibits relatively strong responses (Extended Data Figs. \ref{EDF0}c1). Consequently, the application of an inverse spectral operator whose spatial support matches that of the effective 2D OTF enables targeted suppression of the undesired projected components  (Fig. \ref{fig:principle}d1). In contrast, the first-order spectral components, which are responsible for compensating the missing cone, are associated with projected effective 2D OTFs that exhibit markedly weaker responses (Extended Data Figs. \ref{EDF0}a2-\ref{EDF0}c2). Therefore, the same spectral operation naturally preserves these components to enhance the axial response  (Fig. \ref{fig:principle}d2). As a result, frequency components that compensate for the missing cone are selectively emphasized, whereas components dominated by projected out-of-focus contributions are attenuated. This optimal frequency-weighted reconstruction can be expressed in the frequency domain as:
\begin{equation}
M({\bf{k}}) = \frac{{{C_{mis.}}({{\bf{k}}_{xy}}){\mkern 1mu} {\mkern 1mu} \tilde h_{mis.}^*({{\bf{k}}_{xy}})}}{{{{\left| {{{\tilde h}_{mis.}}({{\bf{k}}_{xy}})} \right|}^2} + {w^2}}} + \frac{{{C_{com.}}({{\bf{k}}_{xy}}){\mkern 1mu} {\mkern 1mu} \tilde h_{com.}^*({{\bf{k}}_{xy}})}}{{{{\left| {{{\tilde h}_{com.}}({{\bf{k}}_{xy}})} \right|}^2} + {w^2}}}
	\label{eq:S6}
\end{equation}
where $C$ and $\tilde{h}$ denote the frequency-shifted spectral components with the modulation depth and initial phase terms removed and their corresponding effective 2D OTFs, respectively, the subscripts $mis.$ and $com.$ indicate the components affected by the missing-cone deficiency (zeroth- and second-order spectra) and those compensating for the missing cone (first-order spectra), respectively, and $w$ is the Wiener constant (regularization parameter) set empirically. Importantly, the complementary projection characteristics of these two classes of spectral components allow the effective 2D OTFs to act as intrinsic weighting functions, enabling physically consistent suppression of out-of-focus contributions while reinforcing axial-frequency information (Supplementary Fig. S2). Unlike empirical fusion strategies that rely on manually tuned frequency weights \cite{johnson2020artifact, wen2021high}, the proposed reconstruction framework determines spectral contributions based on the physical constraints imposed by the 3D OTF, providing improved robustness and generalizability across diverse imaging conditions  (Supplementary Note S5 and Supplementary Figs. S3-S4). Together, this 3D OTF-informed optimal frequency-weighted reconstruction framework completes the 4I-SIM pipeline by fully exploiting the enriched frequency encoding established by four-beam interference and asynchronous phase encoding, enabling intrinsic OS alongside high lateral SR within a standard 2D-SIM acquisition.

\subsection{Numerical simulations to demonstrate the superior comprehensive performance of 4I-SIM in terms of lateral super-resolution, optical sectioning, and reconstruction fidelity}

To evaluate the performance of 4I-SIM, we constructed an ideal 3D two-layer fluorescence model comprising $512 \times 512 \times 9$ voxels (voxel size $65 \times 65 \times 200$ nm$^3$). We positioned a standard high-contrast 2D resolution image in the first layer, defined as the in-focus distribution $\rho_0(x,y)$, to establish the dominant high-frequency information and directional characteristics of the image; simultaneously, we rotated the image by $\pi/2$ and placed it in the ninth layer (corresponding to a $1600$ nm axial shift) to simulate the background distribution $\rho_1(x,y)$. Subsequently, we convolved this ground truth with a 3D point spread function ($\mathrm{NA} = 1.45$, $\lambda = 561$ nm). Since the axial separation of $1.6\,\mu\mathrm{m}$ significantly exceeds the depth of field, the convolution operation exerted a strong low-pass filtering effect on the upper-layer signal, causing it to manifest at the focal plane solely as a low-frequency background with specific directional characteristics, without introducing interfering high-frequency components.

As illustrated in Figs. \ref{fig:simulation}a-\ref{fig:simulation}c, both the conventional SR-SIM and the second-order spectrum reconstruction of 4I-SIM achieve significant lateral resolution enhancement, clearly resolving the denser features near the center of the grating pattern (indicated by blue arrows). However, these reconstructions are adversely affected by defocused background signals, especially in regions highlighted by the orange arrows. In contrast, the first-order spectrum reconstruction provides superior OS capability but limited resolution improvement. By integrating the first- and second-order spectral components, the 4I-SIM reconstruction attains a balance between enhanced lateral resolution and effective suppression of defocused background signals. To quantitatively assess sectioning performance, local contrast ratios were calculated for each method (Fig. \ref{fig:simulation}d). The results demonstrate that 4I-SIM significantly outperforms other methods in this regard. Additionally, Fig. \ref{fig:simulation}e presents lateral resolution distributions obtained via rolling Fourier ring correlation (rFRC) measurements \cite{zhao2023quantitatively}, showing that 4I-SIM achieves resolution improvements comparable to conventional SR-SIM, with mean rFRC resolutions of 104.822 nm and 103.021 nm respectively, corresponding to 1.88$\times$ and 1.90$\times$ resolution enhancement, respectively. To further evaluate reconstruction fidelity, we calculated the resolution-scaled error map (RSM) distribution shown in Fig. \ref{fig:simulation}f, where 4I-SIM demonstrates the best structural fidelity among the compared methods. These results collectively indicate that 4I-SIM achieves superior overall performance by balancing SR and OS, thereby enhancing the reconstruction quality for thick specimen imaging.

\begin{figure}[htbp]
	\centering
	\includegraphics[width=1\textwidth]{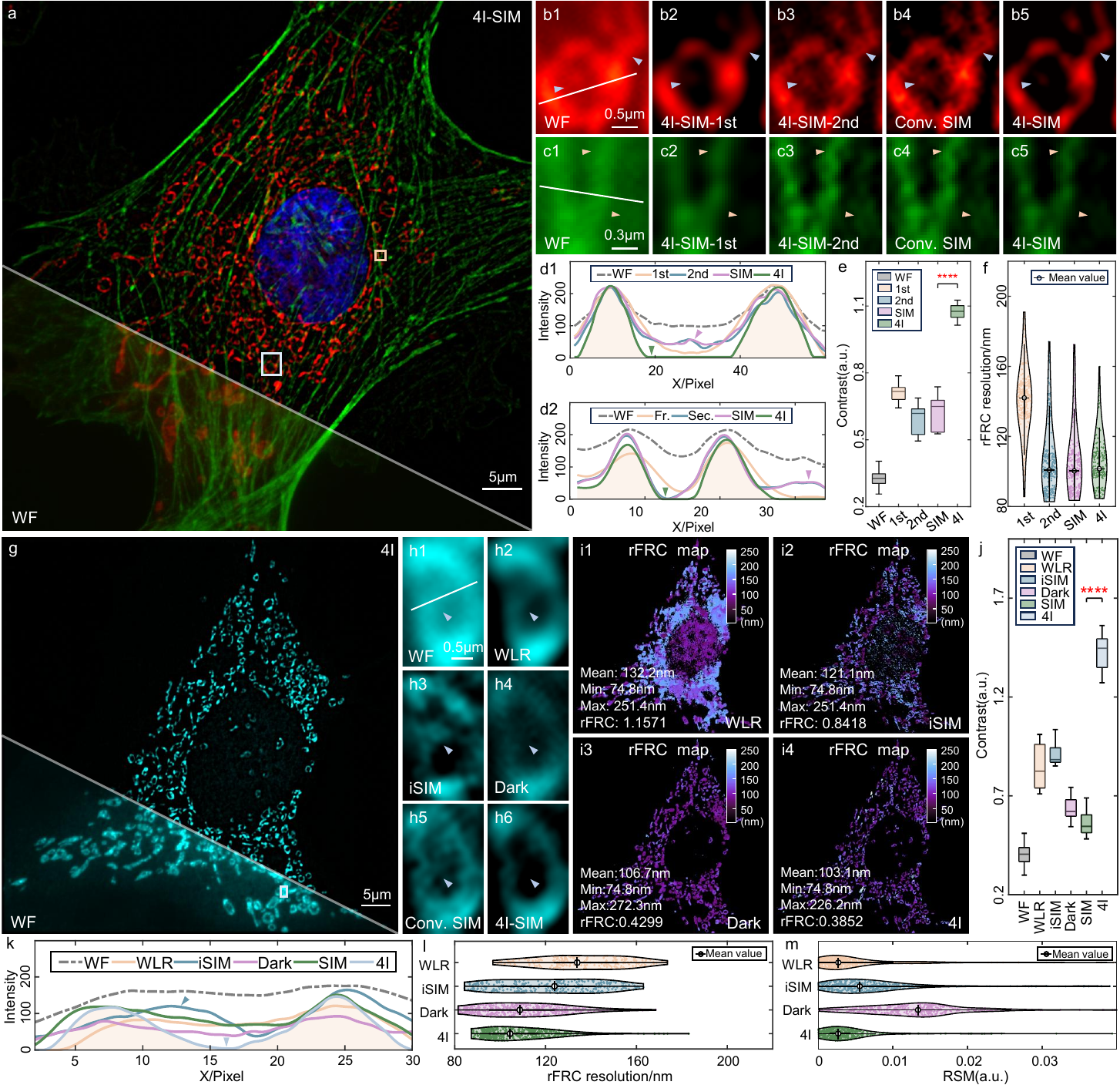}
\end{figure}

\begin{figure}[htb]
	\caption{\textbf{Comparative experiments on fixed BPAE cell samples.} 
		\textbf{a} Comparison between wide-field and 4I-SIM super-resolution images of mitochondria, actin filaments, and nuclei in BPAE cells. The raw images were acquired at $1024 \times 1024$ pixel resolution using a 100$\times$ oil-immersion objective (UPlanXApo 100/1.45 Oil, Olympus, Japan). For easy distinguishing, we show the mitochondria, actin filaments and nucleus in red, green and blue, respectively.  
		\textbf{b, c} Magnified wide-field and super-resolution images from the blue (mitochondria) and yellow (actin filaments) boxed regions in \textbf{a} obtained by different methods (first-order spectrum reconstruction of 4I-SIM, second-order spectrum reconstruction of 4I-SIM, conventional SR-SIM and 4I-SIM).
		\textbf{d} Fluorescence intensity profiles along the white lines in \textbf{b} and \textbf{c}.  
		\textbf{e} Comparison of local image contrast between the wide-field image and super-resolution images obtained by different methods. 
		\textbf{f} Comparison of rFRC lateral resolution distributions between super-resolution images obtained by different methods. 
		\textbf{g} Comparison between wide-field and 4I-SIM super-resolution images of mitochondria in BPAE cells.  
		\textbf{h} Magnified wide-field and super-resolution images from the blue boxed regions in \textbf{g} obtained by different methods (conventional SR-SIM, WLR-SIM \cite{o2014optimized}, iSIM \cite{dan2020super}, Dark-SIM \cite{cao2024dark} and 4I-SIM). 
		\textbf{i} Comparison of rFRC resolution maps between super-resolution images obtained by different methods. 
		\textbf{j} Comparison of local image contrast between the wide-field image and super-resolution images obtained by different methods. 
		\textbf{k} Fluorescence intensity profiles along the white line in \textbf{h}.  
		\textbf{l}  Comparison of rFRC lateral resolution distributions between super-resolution images obtained by different methods. 
		\textbf{m} Comparison of RSM distributions between super-resolution images obtained by different methods.   
		A two-tailed paired Student's $t$-test was applied to contrast values in \textbf{j}, with **** indicating $p < 0.0001$. Each experiment was independently repeated ten times with consistent results. Colored arrows mark regions with significant reconstruction differences. 
		Scale bars: 5 $\mu$m
		(\textbf{a}, \textbf{g}); 500 nm (\textbf{b}, \textbf{h}); 300 nm (\textbf{c}).}
	\label{fig:3color}
\end{figure}

\begin{figure}[htbp]
	\centering
	\includegraphics[width=1\textwidth]{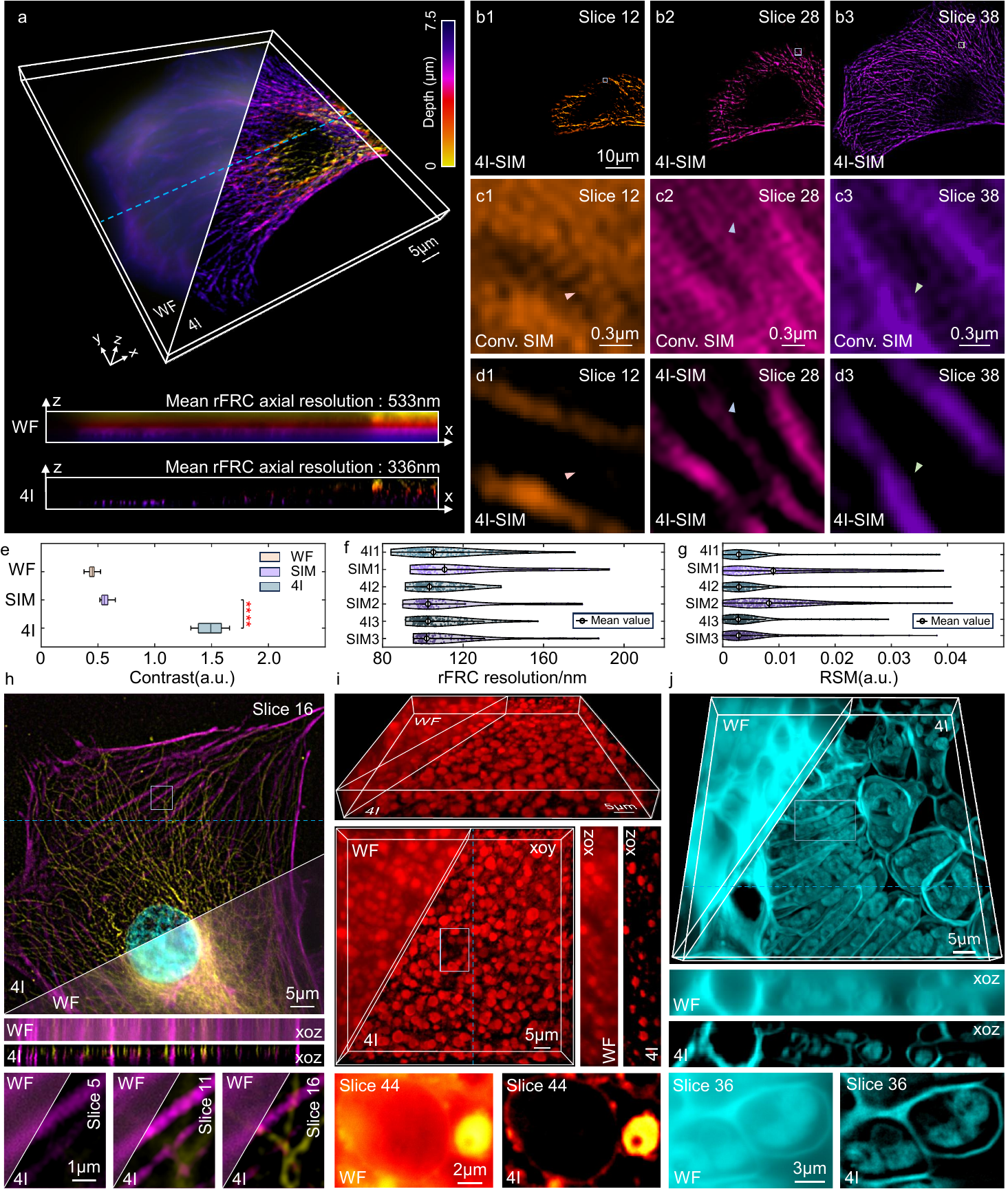}
\end{figure}

\begin{figure}[htb]
	\caption{\textbf{Comparative experiments on volumetric imaging across different thick samples.}  
		\textbf{a} Comparison of wide-field and 4I-SIM super-resolution images of the entire 3D field of view of a fixed BSC-1 cell sample. The bottom panel shows the $x$-$z$ slices of the wide-field and super-resolution images along the blue dashed line. The raw images were captured at a resolution of $1024 \times 1024$ pixels using a 100$\times$ oil-immersion objective (UPlanXApo 100/1.45 Oil, Olympus, Japan).
		\textbf{b} Super-resolution images obtained by 4I-SIM from different axial slices (slices 12, 28, and 38) in \textbf{a}.
		\textbf{c} Magnified super-resolution images from the blue boxed regions in \textbf{b} obtained by conventional SR-SIM.
		\textbf{d} Magnified super-resolution images from the blue boxed regions in \textbf{b} obtained by 4I-SIM.
		\textbf{e} Comparison of local image contrast between the wide-field image and super-resolution images obtained by different methods. 
		\textbf{f} Comparison of rFRC lateral resolution distributions between super-resolution images obtained by different methods. 
		\textbf{g} Comparison of RSM distributions between super-resolution images obtained by different methods. 
		\textbf{h} Comparison between wide-field and 4I-SIM super-resolution images at different axial slices of a fixed COS-7 cell sample. For easy distinguishing, we show the microtubules, actin filaments and nucleus in yellow, magenta, and blue, respectively.
		\textbf{i} Comparison of wide-field and 4I-SIM super-resolution images of the entire 3D field of view of a fixed autofluorescent roundworm sample.
		\textbf{j} Comparison of wide-field and 4I-SIM super-resolution images of the entire 3D field of view of a fixed autofluorescent geranium leaf cross-section.
		In \textbf{f} and \textbf{g}, the numbers following `4I' and `SIM' indicate the corresponding results in \textbf{b$_n$}.
		A two-tailed paired Student's $t$-test was applied to contrast values in \textbf{e}, with **** indicating $p < 0.0001$. All experiments were independently repeated ten times with consistent results. Colored arrows indicate regions with significant differences in image reconstruction.  
		Scale bars: 5 $\mu$m (\textbf{a}, \textbf{h}-\textbf{j}); 10 $\mu$m (\textbf{b}); 300 nm (\textbf{c}, \textbf{d}). Scale on $z$-axis: 50 slices, 150 nm per slice (\textbf{a}); 20 slices, 150 nm per slice (\textbf{h}); 130 slices, 150 nm per slice (\textbf{i}); 70 slices, 150 nm per slice (\textbf{j}).
	}
	\label{fig:multilayer}
\end{figure}

\subsection{Comparative experiments to demonstrate the superior reconstruction capability of 4I-SIM}
To validate the practical performance of 4I-SIM, a custom SIM system using four-beam interference-based composite structured illumination was constructed (see Supplementary Fig. S5 for system details). This platform enabled subsequent systematic comparisons between 4I-SIM and various SR and OS strategies.

We first imaged a fixed BPAE cell sample with the nucleus, actin filaments, and mitochondria labeled by DAPI, Alexa Fluor\textit{\textsuperscript{TM}} 568, and MitoTracker\textit{\textsuperscript{TM}} Green FM, respectively. As shown in Figs. \ref{fig:3color}a-\ref{fig:3color}c, the two enlarged regions of interest exhibit substantial out-of-focus background, which is effectively suppressed by the first-order spectrum reconstruction of 4I-SIM thanks to its compensation for the OTF missing cone. Although the second-order spectrum reconstruction yields higher spatial resolution comparable to conventional SR-SIM, it remains susceptible to defocus-induced artifacts. In contrast, 4I-SIM, integrating all spectral orders, simultaneously eliminates background artifacts and preserves high lateral resolution, as further supported by the fluorescence intensity profiles of mitochondria and actin filaments (Fig. \ref{fig:3color}d). We also evaluated the image contrasts and rFRC resolutions across different methods, as illustrated in Figs. \ref{fig:3color}e and \ref{fig:3color}f. Quantitative contrast analysis (Fig. \ref{fig:3color}e) confirms the superior OS capability of 4I-SIM, which outperforms first-order reconstruction alone after the application of composite frequency-domain filtering that fully exploits spectral components with strong axial responses. Regarding resolution, the rFRC resolution distributions in Fig. \ref{fig:3color}f indicate that 4I-SIM achieves a lateral resolution of 100.876 nm, which is slightly lower than that of conventional SR-SIM (101.334 nm). Additional comparative experiments on BPAE cells with different target configurations further confirmed that 4I-SIM yields superior SR image quality (Supplementary Figs. S6a and S6b). We also extended the comparison to other samples, including a fixed autofluorescent roundworm sample collected with our experimental setup (Supplementary Figs. S6c and S6d) and microtubules in COS-7 cells (Extended Data Fig. \ref{EDF1}). Across all these datasets, 4I-SIM consistently delivered higher-quality SR reconstructions with enhanced OS.

We subsequently compared 4I-SIM with other SIM variants optimized for thick-sample imaging, including WLR-SIM \cite{o2014optimized}, iSIM \cite{dan2020super}, and Dark-SIM \cite{cao2024dark}. The reconstruction results obtained using these techniques are shown in Figs. \ref{fig:3color}g-\ref{fig:3color}m. While all methods exhibit varying degrees of SR and OS capabilities, each suffers from specific limitations—WLR-SIM shows constrained resolution, iSIM may introduce slight fidelity loss, and Dark-SIM enhances contrast but overlooks finer structural features. In contrast, 4I-SIM reveals mitochondrial features with sharper boundaries and reduced background (Fig. \ref{fig:3color}h), and exhibits a more uniform and refined resolution distribution across the field, as visually observed in the rFRC maps (Fig. \ref{fig:3color}i). Further quantitative evaluations, including image contrast, rFRC resolution, and RSM distribution (Figs. \ref{fig:3color}j-\ref{fig:3color}m), demonstrate that 4I-SIM delivers  superior overall performance  in terms of spatial resolution, OS strength, and reconstruction fidelity.

\subsection{Experiments to demonstrate the volumetric imaging capability of 4I-SIM in thick biological specimens}
The excellent OS and SR capabilities of 4I-SIM make it well-suited for volumetric imaging of thick biological samples, which can be readily achieved by incorporating axial scanning  into the imaging process. To validate this, we integrated a nanometer-precision $z$-axis translation stage into our custom-built imaging system and performed sequential layer-by-layer scanning of a thick BSC-1 cell sample (with Alexa Fluor$^{TM}$ 555-labeled microtubules). At each axial position, 4I-SIM reconstruction was carried out accordingly to generate a volumetric result of the sample. As shown in Fig. \ref{fig:multilayer}a, 4I-SIM resolves microtubule networks across the entire 3D volume with clear structural delineation across depth. The bottom panel compares the $x-z$ slices of wide-field  and 4I-SIM results along the blue dashed line, revealing significantly enhanced axial resolution and reduced background in the 4I-SIM image. Based on the rFRC method, the mean rFRC axial resolution of 4I-SIM was measured to be 336.2 nm, representing a 1.59-fold improvement over the 534.6 nm resolution achieved by wide-field imaging. To assess performance across depth, we analyzed three specific axial planes (Slices 12, 28, and 38, as illustrated in Figs. \ref{fig:multilayer}b-\ref{fig:multilayer}d). 4I-SIM consistently reveals fine microtubule filaments with high clarity, while conventional SR-SIM exhibits elevated background and defocus-induced reconstruction artifacts. Quantitative comparisons of image contrast (Fig. \ref{fig:multilayer}e), lateral resolution (Fig. \ref{fig:multilayer}f), and reconstruction fidelity (Fig. \ref{fig:multilayer}g) further confirm the advantages of 4I-SIM, which achieves higher contrast, finer resolution (with a mean rFRC lateral resolution of 103.4 nm, approximately a 1.91-fold improvement over wide-field imaging), and better reconstruction consistency (lower RSM) across all depths. Furthermore, we applied 4I-SIM to additional thick biological specimens, including COS-7 cell sample (Fig. \ref{fig:multilayer}h), autofluorescent roundworm sample (Fig. \ref{fig:multilayer}i), autofluorescent geranium leaf cross-section (Fig. \ref{fig:multilayer}j), autofluorescent ascaris suum fertilized eggs sample (Extended Data Fig. \ref{EDF2}a), chlamydomonas sample (Extended Data Fig. \ref{EDF2}b), mouse brain section (Extended Data Fig. \ref{EDF2}c), and mouse kidney section (Extended Data Figs. \ref{EDF2}d). Supplementary Videos S2-S8 show 3D volumetric renderings of these specimens. These results collectively confirm that 4I-SIM enables robust and high-quality volumetric imaging across a wide range of thick biological specimens.

\begin{figure}[htb]
	\centering
	\includegraphics[width=1\textwidth]{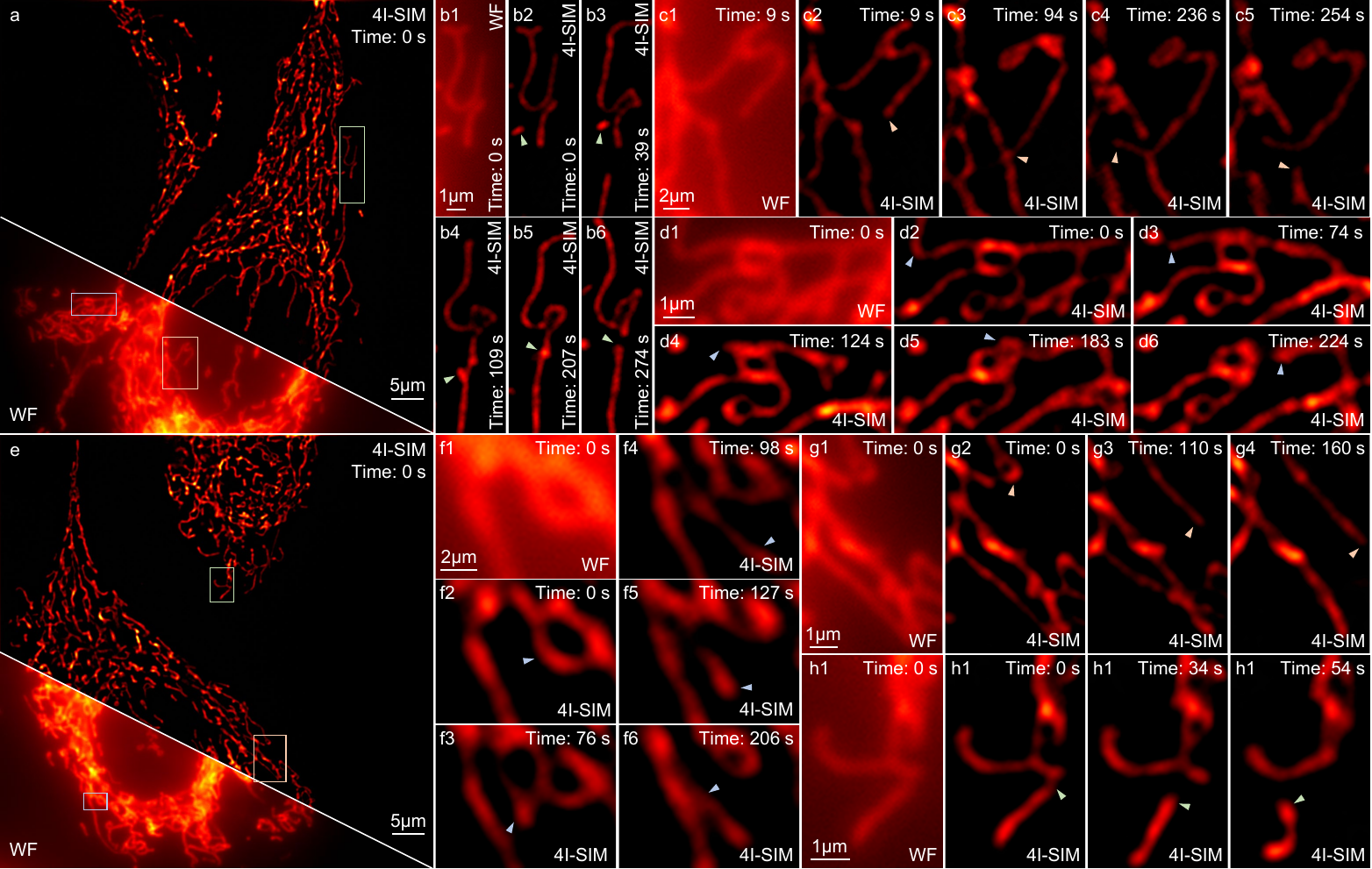}
	\caption{\textbf{Super-resolution reconstruction results of mitochondria in live HeLa cells at different time points.}  
		\textbf{a}, \textbf{e} Comparison of wide-field and 4I-SIM reconstructed super-resolution images. The raw images were acquired at $1024 \times 1024$ pixel resolution using a 100$\times$ oil-immersion objective (UPlanXApo 100/1.45 Oil, Olympus, Japan).  
		\textbf{b}, \textbf{h} Time-lapse super-resolution images of the green boxed regions in \textbf{a} and \textbf{e}, respectively.  
		\textbf{c}, \textbf{g} Time-lapse super-resolution images of the yellow boxed regions in \textbf{a} and \textbf{e}, respectively. 
		\textbf{d}, \textbf{f} Time-lapse super-resolution images of the blue boxed regions in \textbf{a} and \textbf{e}, respectively. 
		Scale bars: 5 $\mu$m
		(\textbf{a}, \textbf{e}); 1 $\mu$m
		(\textbf{b}, \textbf{d}, \textbf{g}, \textbf{h}); 2 $\mu$m
		(\textbf{c}, \textbf{f}).}
	\label{fig:livecell}
\end{figure}

\subsection{Dynamic super-resolution imaging of mitochondria in live HeLa cells under complex biological conditions using 4I-SIM}

Unlike fixed specimens, live-cell imaging imposes substantially greater demands on OS and reconstruction fidelity due to complex 3D architectures, imperfect labeling, and dynamically varying background scattering \cite{schneckenburger2021challenges,frigault2009live,wang2024high, nath2012identification,han2017cell,kollmannsperger2016live}. Under these conditions, conventional SIM often suffers from severe defocus contamination and background accumulation, leading to unstable reconstructions and unreliable interpretation of dynamic subcellular processes, as seen in Supplementary Fig. S7. In this context, 4I-SIM provides a robust imaging solution that maintains OS and SR performance under complex live-cell conditions, without relying on optimized labeling density or background suppression heuristics, as demonstrated by the experimental results acquired with 4I-SIM on mitochondrial dynamics in live HeLa cells labeled with PK Mito Red, where distinct mitochondrial behaviors were clearly observed (Fig. \ref{fig:livecell}). As illustrated in Figs. \ref{fig:livecell}a and \ref{fig:livecell}b, a highly motile mitochondrion (green arrow) was observed migrating between two elongated, parallel-aligned mitochondria (0 s-109 s), where it rapidly fused upon contact. It was then fully incorporated into the fused structure, forming a continuous mitochondrial filament (207 s). At 274 s, a distinct fission event occurred at the original fusion site, indicating a dynamic remodeling process in which motile mitochondria contribute to network fusion and subsequently undergo regulated fission. Figure \ref{fig:livecell}c provides a clear visualization of another mitochondrial remodeling event. At 9 s, the mitochondrion indicated by the pink arrowhead appeared as an isolated structure. At 94 s, it gradually extended toward and fused with the adjacent mitochondrial network. Shortly thereafter, between 236 s and 254 s, a fission event occurred at the fusion site, leaving the mitochondrion again as an isolated structure. Figure \ref{fig:livecell}d presents additional representative remodeling and fission events. At 0 s, the indicated mitochondrion (blue arrow) appeared as an extended tubular structure. Between 74 s and 183 s, it gradually retracted and merged into the main network, leaving behind a detached punctate structure. At 224 s, a visible break occurred at the narrow junction between a ring-like structure and the main network, suggesting that mitochondrial fission preferentially occurs at structurally constrained or transitional regions. Figures \ref{fig:livecell}e-\ref{fig:livecell}h highlights diverse mitochondrial remodeling behaviors. In one instance, a ring-shaped mitochondrion gradually extended a tubular projection that detached from the parent and ultimately fused with a neighboring network (Fig. \ref{fig:livecell}f). In another, a mitochondrion initially exhibited a coiled morphology, which progressively straightened into an elongated, extending structure (Fig. \ref{fig:livecell}g). Additionally, a fission event occurred at a network junction, giving rise to an isolated fragment that subsequently adopted a constricted, curled shape (Fig. \ref{fig:livecell}h). These dynamic processes reveal the intricate regulation of mitochondrial morphology associated with metabolic adaptation, quality control, and structural reorganization \cite{youle2012mitochondrial,sarraf2013landscape,jenkins2024mitochondria,roca2021mitochondrial,chen2023mitochondrial}. Since 4I-SIM requires the same number of raw images as conventional 2D-SIM and is compatible with PCA-based parameter estimation, it achieves a reconstruction efficiency comparable to PCA-SIM \cite{qian2023structured}, enabling real-time imaging at $\sim$30 Hz. The complete time-lapse videos are provided in Supplementary Videos S9 and S10. Together, these results demonstrate that 4I-SIM enables reliable SR imaging of dynamic mitochondrial remodeling in live cells under conditions where conventional SIM fails due to insufficient OS.

\begin{figure}[htbp]
	\centering
	
	\includegraphics[width=0.9\textwidth]{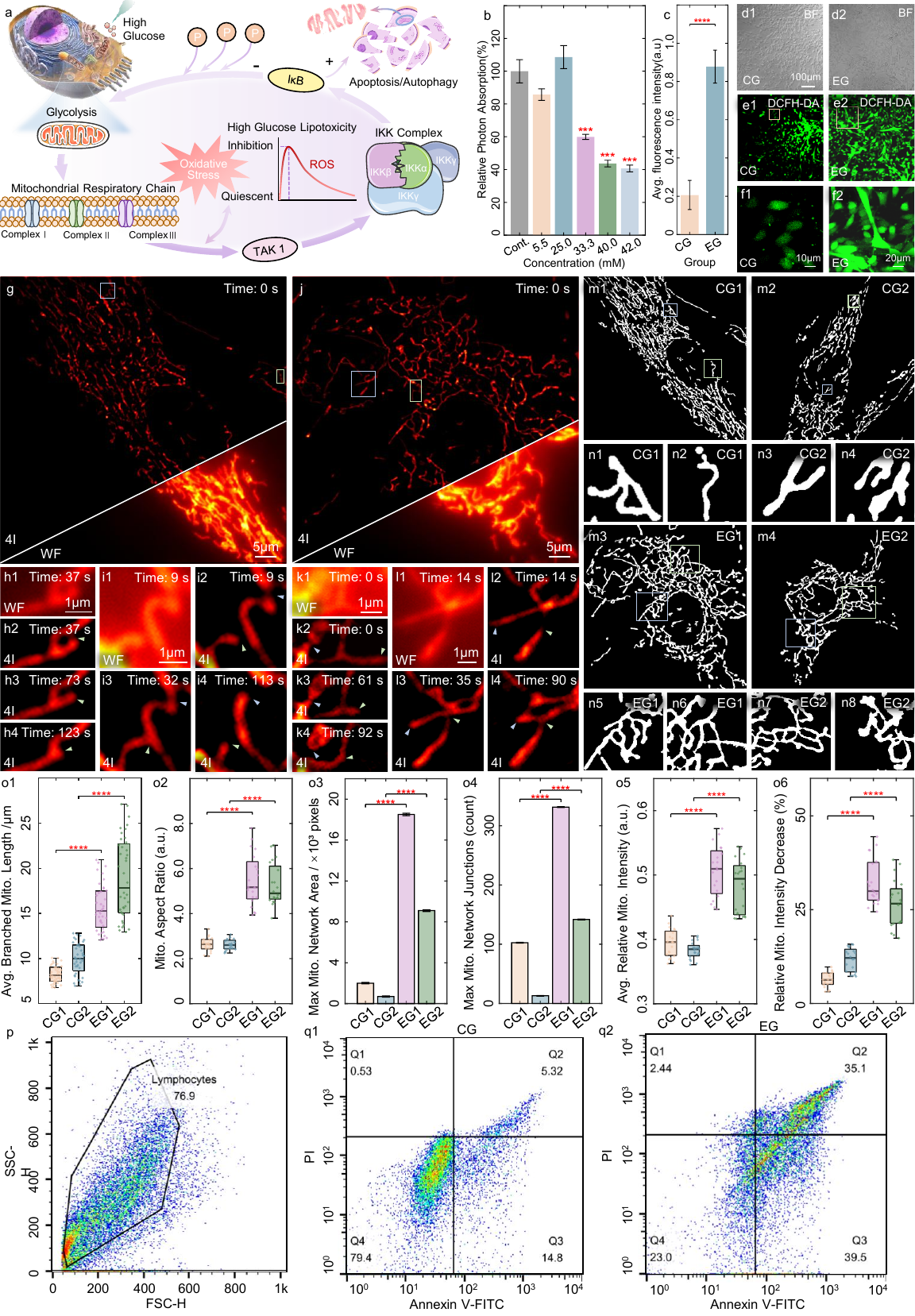}
\end{figure}

\begin{figure}[htb]
	\centering
	\caption{\textbf{Super-resolution imaging analysis of mitochondria in live human myofibroblasts under high-glucose conditions.}  
		\textbf{a} Response mechanism of oxidative stress and NF-$\kappa$B signaling pathway under high glucose environment.
		\textbf{b} High-glucose concentration was assayed by CCK-8 kit.
		\textbf{c} Average fluorescence intensity of the DCFH-DA probe at different groups.
		\textbf{d} Cell morphology at different groups was detected by bright-field microscopy (20$\times$).
		\textbf{e} The expression of ROS was imaged by confocal microscopy (20$\times$).
		\textbf{f} Magnified images of the regions marked with boxes in \textbf{e}.
		\textbf{g}, \textbf{j}  Comparison of widefield and 4I-SIM reconstructed super-resolution images under normal glucose conditions (control group) and high glucose conditions (experimental group), respectively.
		Mitochondria were labeled with PK Mito Red and excited using a 561 nm laser. Raw SIM data were acquired at $1024 \times 1024$ pixel resolution using a 100$\times$ oil-immersion objective (UPlanXApo 100/1.45 Oil, Olympus, Japan).
		\textbf{h}, \textbf{k}  Time-lapse super-resolution images of the green boxed region in \textbf{g} and \textbf{j}, respectively. 
		\textbf{i}, \textbf{l}  Time-lapse super-resolution images of the blue boxed region in \textbf{g} and \textbf{j}, respectively.  
		\textbf{m}  Representative binary images of mitochondria in control (normal glucose; m1, m2) and experimental (high glucose; m3, m4) groups.
		\textbf{n}  Magnified view of the boxed region in \textbf{m}. 
		\textbf{o}  Comparative analysis of multiple mitochondrial parameters in control (normal glucose) and experimental (high glucose) groups.
		\textbf{p} Cell population dot plot. 
		\textbf{q} Four-quadrant plots  of the Control Group (q1) and the Experimental Group (q2).
		A two-tailed paired Student's $t$-test was used to compare values in \textbf{b}, \textbf{c}, and \textbf{o}; *** and **** indicate $p < 0.001$ and $p < 0.0001$, respectively.All experiments were independently repeated ten times with consistent results. Colored arrows indicate regions with significant differences in image reconstruction.  
		Scale bars: 100 $\mu$m (\textbf{d}, \textbf{e}); 10 $\mu$m (\textbf{f}); 5 $\mu$m (\textbf{g}, \textbf{j}); 1 $\mu$m (\textbf{h}, \textbf{i}, \textbf{k}, \textbf{l}).
	}
	\label{fig:highglucose}
\end{figure}

\subsection{Investigating mitochondrial structure and function under high-glucose conditions using 4I-SIM}

Beyond conventional mitochondrial dynamics imaging, 4I-SIM is also capable of tackling more complex biological scenarios. One such scenario involves investigating mitochondrial responses under prolonged high-glucose conditions, which significantly affect cellular physiology. As the central hub of cellular energy and metabolism, mitochondria have been regarded as key targets for high-glucose-induced damage \cite{yu2006increased,otero2025hyperglycemia,jia2024epigallocatechin,aluksanasuwan2020high,xu2024high,wu2021high}. Under such pathological states, the level of reactive oxygen species derived from mitochondria (mtROS) is significantly elevated, the results of which oxidative stress responses and apoptosis factors are activated dramatically. We conducted the response of oxidative stress and the classical signaling pathway in high-glucose environments (Fig. \ref{fig:highglucose}a). Lipotoxicity is initiated when glycolysis proceeds under high-glucose conditions and the mitochondrial respiratory-chain complexes are drastically remodeled. The explosive micro-molecule, ROS and inflammatory mediators, activates the NF-$\kappa$B pathway. The upstream regulators such as TAK1 and the IKK complex are engaged, provoking cellular injury and fragmentation of the mitochondrial network. Phosphorylation of the downstream target I$\kappa$B can transiently restrain this damage, but once the homeostatic balance is lost the cell is funneled into an irreversible fate such as apoptosis or autophagy.

To investigate these mitochondria mechanisms under high-glucose conditions, we employed 4I-SIM to dynamically monitor mitochondrial structure and function changes under high-glucose conditions in vivo, providing new insights into the pathophysiology of metabolic disorders. Firstly, cell viability was detected by the CCK-8 kit, which showed a marked reduction under the high-glucose conditions compared to normal glucose (25 mM) conditions, with relative absorbance reduced to 0.64 (Fig. \ref{fig:highglucose}b). 
The oxidative stress was subsequently assessed via intracellular ROS, as shown in Fig. \ref{fig:highglucose}c. 
In this study, cells cultured under normal glucose concentration were designated as the control group (CG), while those exposed to high-glucose conditions were defined as the experimental group (EG).
Notably, high-glucose conditions impaired cell adhesion and thickness, leading to elevated background fluorescence, increased scattering, and pronounced structural heterogeneity, which collectively exacerbated the cellular microenvironment, increased imaging complexity, and heightened background noise.
Semi-quantitative confocal analysis further confirmed that the ROS accumulation promoted mitochondrial dysfunction and apoptosis (Figs. \ref{fig:highglucose}d-\ref{fig:highglucose}f). 
These conditions pose a fundamental challenge to conventional SIM, where insufficient os and background accumulation severely compromise reconstruction fidelity and long-term imaging stability (Supplementary Fig. S8).
In contrast, 4I-SIM preserves both lateral SR and robust OS under high-glucose conditions, enabling high-fidelity, long-term live imaging of mitochondrial structure and network remodeling in optically challenging environments. (Figs. \ref{fig:highglucose}g-\ref{fig:highglucose}l, and Supplementary Videos S11-S12). To quantify mitochondrial alterations, we analyzed multiple morphological and functional parameters, including mitochondrial length, aspect ratio, maximal network area, counts of network nodes, as well as fluorescence intensity and its decay rate (Figs. \ref{fig:highglucose}m-\ref{fig:highglucose}o). These metrics comprehensively reflect the structural remodeling capacity of mitochondria, network-level spatial organization and connectivity, and the dynamic changes in metabolic homeostasis and functional integrity. Under high-glucose conditions, mitochondria exhibited disrupted spatial orientation and diminished structural order. Aspect ratio analysis showed significant mitochondrial shrinkage and reduced matrix density. Additionally, analysis of network fusion confirmed that mitochondrial network integrity was severe disruption, thereby accelerating cellular apoptosis. Intriguingly, 4I-SIM also captured subtle changes in mitochondrial membrane morphology, such as blurring and rupture, which indirectly verified that cellular photostability was partially impaired (113 s vs. 90 s). Previous studies have identified calcium ions as key regulators of mitochondrial membrane stability and oxidative stress \cite{borbolis2025calcium, cheng2003vdac2, shimizu1999bcl, zhang2025restoring, perez2016optimal}. We thus further investigated calcium ion accumulation under high-glucose conditions and its impact on mitochondrial function. Confocal imaging revealed excessive intracellular calcium, which increased mitochondrial membrane permeability and led to the loss of essential proteins like cytochrome c.
Flow cytometry also confirmed that the high-glucose conditions induced premature apoptosis, as shown as Figs. \ref{fig:highglucose}p-\ref{fig:highglucose}q. The cells in CG were concentrated in the quadrant IV (Annexin V- and FITC-PI-), which was normal cells. The cells in EG were concentrated in the quadrant II and III, which indicated that early apoptosis (39.5$\%$) and late apoptosis (35.1$\%$) events was happened simultaneously.

In summary, high-glucose conditions significantly impaired cell viability and disrupted mitochondrial homeostasis, partly via excessive calcium ion accumulation and increased oxidative stress. These pathological changes ultimately led to apoptosis. The robust imaging capability of 4I-SIM under such challenging conditions enabled detailed structural and functional mitochondrial analyses at the SR level, offering valuable insights into the cellular mechanisms of oxidative damage in metabolic disease contexts.

\section{Discussion}\label{sec3}

In summary, we have introduced four-beam interference structured illumination microscopy (4I-SIM), a 2D SIM-based framework that achieves robust OS while preserving the lateral SR and acquisition efficiency characteristic of conventional SR-SIM. By explicitly reinterpreting 2D SIM from a 3D transfer-function perspective, 4I-SIM addresses the long-standing axial missing-cone limitation through engineered frequency encoding and physically informed reconstruction, enabling high-fidelity imaging in thick and optically complex biological specimens.
A central advantage of 4I-SIM lies in its ability to integrate the complementary strengths of 2D- and 3D-SIM without inheriting their respective limitations. Specifically, 4I-SIM retains the imaging speed, low phototoxicity, and hardware simplicity of 2D SIM, while substantially improving os and suppressing defocus- and scattering-induced artifacts that typically compromise image fidelity in thick samples. Unlike volumetric SIM approaches, this enhancement is achieved without axial scanning, increased acquisition burden, or specialized optical configurations, ensuring seamless compatibility with existing SIM platforms. From a methodological standpoint, 4I-SIM differs fundamentally from prior efforts that combine SR and OS through empirical background suppression or manually tuned spectral fusion \cite{johnson2020artifact, wen2021high, cao2024dark}. The proposed reconstruction framework is guided by the physical constraints of the 3D OTF, enabling optimal frequency weighting based on intrinsic axial sensitivity rather than specimen-specific heuristics. This physics-informed strategy provides stable and reproducible performance across diverse imaging conditions, reducing dependence on expert parameter tuning and enhancing generalizability for practical biological applications.

It is important to emphasize, however, that while 4I-SIM effectively compensates for the axial missing-cone region of the system OTF, it does not extend the detectable axial spatial-frequency support beyond the intrinsic cutoff imposed by the optical system. As such, 4I-SIM does not achieve genuine axial SR. In fact, the resolution limit of a microscopy modality is fundamentally limited by its signal-to-noise ratio (SNR) \cite{cox1986information, wang2019deep}. From an information-theoretic perspective, the apparent improvement in axial resolution primarily reflects an enhancement of the effective SNR through suppression of out-of-focus background, resulting in improved axial contrast rather than an increase in true axial resolving power. Accordingly, the quantitative resolution improvements observed in this work should be interpreted in terms of enhanced OS and structural fidelity. Regarding lateral resolution, 4I-SIM extends the accessible spatial-frequency range by exploiting second-order spectral components. Although this extension is sufficient to achieve near twofold lateral SR, the resulting frequency coverage is inherently less uniform than that of conventional SIM. While this non-uniformity does not manifest as a noticeable degradation in rFRC-based resolution metrics, it represents a potential limitation in the frequency domain. Future implementations may mitigate this effect through strategies such as alternating or rotational illumination schemes \cite{sun2018single, fan2019single}, which could further homogenize lateral frequency support without increasing acquisition complexity. More broadly, 4I-SIM highlights a promising direction for advancing optical microscopy through 3D frequency engineering within 2D acquisition frameworks. By decoupling axial discrimination from volumetric scanning, this approach opens new opportunities for developing high-throughput, low-phototoxicity imaging methods that remain compatible with live-cell and long-term observations. Future efforts aimed at extending this paradigm toward true axial SR or integrating adaptive frequency encoding strategies may further expand the capabilities of SIM-based imaging.
To facilitate broad adoption, the open-source implementation of 4I-SIM, together with representative datasets, has been made publicly available (Supplementary Note S7). We anticipate that 4I-SIM will serve not only as a practical tool for high-fidelity SR imaging in complex biological specimens, but also as a conceptual framework for rethinking how 3D information can be optimally encoded and exploited within fundamentally 2D imaging modalities.

\clearpage

\section{Methods}\label{sec4}
\subsection{SIM setup}
The tricolor composite SIM system developed in this study is illustrated in Supplementary Fig. S5. To achieve efficient coupling of multi-wavelength laser sources, a series of dichroic mirrors (DM1: ZT561dcrb, Chroma, USA; DM2: ZT488dcrb, Chroma, USA) and planar mirrors (M1, M2: OMM1-A1, JCOPTiX, China) were employed to sequentially combine three laser lines into a single optical path (Laser 1: OBIS LX405, Coherent, USA; Laser 2: OBIS LX561, Coherent, USA; Laser 3: Sapphire 488LP-200, Coherent, USA). The combined beam then passes through a spatial filter and an achromatic lens (L1: LSB08-A, 150 mm, Thorlabs, USA) for beam shaping, expansion, and collimation, yielding a stable and uniform incident beam suitable for downstream modulation and structured illumination. A half-wave plate (HW1: GCL-0604, Daheng Optics, China) converts the beam to p-polarization to minimize energy loss in the subsequent optical path. The beam is then directed by a polarizing beam splitter (PBS: PBS251, Thorlabs, USA) onto a ferroelectric liquid crystal spatial light modulator (SLM: QXGA-3DM, Fourth Dimension Displays, UK), where a composite grating pattern is displayed. The diffracted beam from the SLM is converted to s-polarization using a second half-wave plate (HW2: GCL-0604, Daheng Optics, China) and further reflected and collimated by another PBS and an achromatic lens (L2: LSB08-A, 250 mm, Thorlabs, USA). At the back focal plane (\emph{i.e.}, the Fourier plane), a precision-designed mask is placed to block the 0th-order diffraction and retain only the $\pm$1st-order components along two orthogonal directions, thereby generating four mutually coherent high-order beams. To achieve balanced modulation and enhance the contrast of the interference fringes, a third half-wave plate (HW3: GCL-0604, Daheng Optics, China) was employed for polarization optimization. The polarization direction of this plate was precisely aligned with the bisector of the polarization directions of the two orthogonal coherent beams. The beams were then sequentially focused onto the back focal plane of the objective using two achromatic lenses (L3: 200 mm; L4: 175 mm, both LSB08-A series, Thorlabs, USA), resulting in high-contrast structured illumination at the sample plane. Fluorescence emitted from the sample is collected by the same objective and passes through a filter set (FM: U-FVN, U-FBW, U-FYW, Olympus, Japan), a tube lens (TL), and is ultimately detected by an sCMOS camera (PCO Edge 5.5, PCO, Germany) with a quantum efficiency of 60$\%$. To ensure temporal synchronization between the SLM and image acquisition, camera exposure is precisely triggered by the SLM controller. Unlike conventional SIM systems employing three illumination directions and three-step phase shifting based on two-beam interference, the proposed system adopts a four-beam interference strategy with nine-step asynchronous phase modulation. Specifically, the SLM displays a composite grating formed by the superposition of two orthogonal sinusoidal patterns. During imaging, this pattern is sequentially encoded with nine asynchronous phase steps to achieve efficient structured illumination.

\subsection{Image reconstruction}

In 4I-SIM, the composite illumination pattern can be expressed as:
\begin{align}
	I(\mathbf{r}) =\ & 1 + m^{1}_u \cdot \cos\big[2\pi\, (\mathbf{k}_u - \mathbf{k}_v) \cdot \mathbf{r} + (\varphi_u - \varphi_v) + \varphi^{1}_u\big] \notag \\ 
	&+ m^{1}_v \cdot \cos\big[2\pi\, (\mathbf{k}_u + \mathbf{k}_v) \cdot \mathbf{r} + (\varphi_m + \varphi_n - \pi) + \varphi^{1}_v\big] \notag \\
	&+ m^{2}_u \cdot \cos\big[4\pi\, \mathbf{k}_u \cdot \mathbf{r} + (2\varphi_m - \pi) + \varphi^{2}_u\big] \notag \\
	&+ m^{2}_v \cdot \cos\big[4\pi\, \mathbf{k}_v \cdot \mathbf{r} + (2\varphi_n - \pi) + \varphi^{2}_v\big] &
	\label{eq3}
\end{align}
where \( \mathbf{r} \) is the spatial coordinate.
Notably, we did not adopt the conventional nine-step synchronous phase-shifted scheme. When the number of phase-shifted steps is increased, the tolerance to intrinsic phase errors of SLM is significantly reduced. This effect becomes more severe in complex illumination scenarios involving nine spectral components, as encountered in 4I-SIM. To address this issue, we employed a nine-step \emph{asynchronous} phase-shifting strategy. In this scheme, $\mathbf{k}_u$ and $\mathbf{k}_v$ in Eq. \ref{eq3} are shifted independently in time, with each executing a three-step phase-shifted sequence (see Supplementary Table S1).
In the image processing pipeline, raw SIM images were first subjected to mild edge tapering to suppress boundary artifacts \cite{gustafsson2008three}. To enhance the signal-to-noise ratio and mitigate Poisson noise, preliminary restoration was performed using the Richardson-Lucy (RL) deconvolution algorithm \cite{perez2016optimal,sage2017deconvolutionlab2}. During illumination parameter estimation, a nine-step phase-shifted matrix was applied to preliminarily separate the raw images into nine modulated spectral components. A PCA-based algorithm was then employed to extract the illumination wavevectors and phases, enabling fast, non-iterative, and accurate parameter estimation \cite{qian2023structured} (detailed in Supplementary Note S4). With these parameters, the nine spectral components were precisely demodulated using the same nine-step matrix. For spectral reconstruction, the zeroth-order and four second-order components were combined to form the high-frequency second-order spectrum. Similarly, four first-order components were used to construct the low-frequency first-order spectrum. Finally, the complete 4I-SIM spectrum was reconstructed through the 3D OTF-informed optimal frequency-weighting reconstruction framework (Eq. \ref{eq:S6}), integrating both spectral channels to achieve twofold lateral SR and effective OS. For benchmarking, alternative reconstruction methods, including conventional SR-SIM \cite{gustafsson2000surpassing}, WLR \cite{o2014optimized}, iSIM \cite{dan2020super}, and Dark-SIM \cite{cao2024dark}, were implemented according to their original protocols to ensure fair and consistent comparison.

\subsection{Sample preparation}
HeLa cells were incubated in 90\% RPMI-1640, 10\% FBS and 1\% penicillin-streptomycin. Biological sample was prepared when cells are exponentially proliferating. Cells were seeded at a density of $5 \times 10^8$ cells in the dishes, which were cultured overnight. The cells were stained through immunofluorescence staining with the cell density reaching 60\%. PK Mito Red ($\lambda_\text{abs}=549$ nm; $\lambda_\text{fl}=569$ nm) were diluted to dimethyl sulfoxide (DMSO) at the concentration of 250 $\mu$M. The primary liquor was diluted by the pre-warmed medium. They were washed by 0.5 mL phosphate buffer (PBS) for 5 min. The prepared solution was incubated for 15 min. To remove interference from free dyes, all samples were washed by pre-warmed PBS for 3 times. CCC-ESF-1 were cultured in vitro in DMEM medium supplemented with 10\% fetal bovine serum and incubated at 37 $^\circ$C with 5\% CO$_2$. Cells incubated for 48 h in DMEM containing 5.5 mmol/L glucose were regarded as the control group, whereas those incubated for 48 h in DMEM containing 33.3 mmol/L glucose were regarded as the high-glucose group.

\subsection{Statistical analysis}
Except for Fig. \ref{fig:livecell} and Fig. \ref{fig:highglucose}, all presented results represent typical outcomes selected from ten independently repeated and representative experiments. In Figs. \ref{fig:simulation}-\ref{fig:multilayer} and Extended Data Fig. \ref{EDF1}, local image contrast comparisons across different methods are presented as box plots, where the central line denotes the mean, the box edges correspond to the 75th and 25th percentiles, and the whiskers indicate the maximum and minimum values. The rFRC-based resolution distributions (Figs. \ref{fig:simulation}-\ref{fig:multilayer} and Extended Data Fig. \ref{EDF1}) and RSM distributions (Figs. \ref{fig:simulation}-\ref{fig:multilayer}) were calculated using the algorithm described in Ref. \cite{zhao2023quantitatively} and visualized as violin plots. These violin plots are rendered symmetrically: the contour represents the kernel density estimation, the width reflects the distribution concentration, the horizontal line marks the mean, the box edges denote the 75th and 25th percentiles, and the whiskers show the data range from minimum to maximum. Additionally, the rFRC resolution maps shown in Fig. \ref{fig:multilayer} were computed using the same method from Ref. \cite{zhao2023quantitatively}. All intensity profile curves presented in Figs. \ref{fig:simulation}-\ref{fig:multilayer} were generated using linear interpolation in MATLAB. Quantitative analysis of various mitochondrial morphological and functional parameters in Fig. \ref{fig:highglucose}, including mitochondrial length, aspect ratio, network area, counts of network nodes, and fluorescence intensity, was performed using Fiji \cite{schindelin2012fiji} with the MiNA (Mitochondrial Network Analysis) plugin \cite{valente2017simple} and related analytical tools. 

The specific definition of the rFRC measurements is given as follows. Local image resolution was quantified using rolling Fourier ring correlation. For each sliding window of size $w \times w$ pixels centered at position $(x,y)$, two statistically independent reconstructions of the same structure, denoted $M_1$ and $M_2$, were transformed into the Fourier domain, and the Fourier ring correlation as a function of spatial frequency $q_i$ was defined as:
\begin{equation}
        \mathrm{FRC}_{12}(q_i)
        =
        \frac{\displaystyle\sum_{r\in q_i} F_1(r)\,F_2^*(r)}
        {\sqrt{\displaystyle\sum_{r\in q_i}\lvert F_1(r)\rvert^2
                        \;\displaystyle\sum_{r\in q_i}\lvert F_2(r)\rvert^2}} 
        \label{eq:frc}
\end{equation}
where $F_1$ and $F_2$ denote the discrete Fourier transforms of $M_1$ and $M_2$, $r$ is the 2 spatial-frequency vector, and the summation runs over all Fourier samples lying on the ring with radius $q_i$. Spatial frequencies were grouped into concentric rings between zero and the Nyquist frequency, and the raw FRC curve was smoothed along the frequency axis using a short moving-average filter to suppress statistical fluctuations. For each ring, the local cut-off frequency $f_{\mathrm{co}}(x,y)$ was defined as the first spatial frequency at which $\mathrm{FRC}_{12}(q_i)$ dropped below a sigma-factor threshold $\sigma_i = 3/\sqrt{N_i/2}$, where $N_i$ represents the number of Fourier samples in the ring. The rFRC resolution map $FV(x,y)$ was then obtained as the reciprocal of this cut-off frequency. To summarize the overall image quality with a single number, the global rFRC resolution was computed as the arithmetic mean of the local rFRC resolution over all foreground pixels:
\begin{equation}
        \mathrm{Mean\ Resolution} =
        \frac{\displaystyle\sum_{FV(x,y)\neq 0} FV(x,y)}
        {\bigl\|FV(x,y)\bigr\|_0} 
        \label{eq:mean_resolution}
\end{equation}
where \(\bigl\|FV(x,y)\bigr\|_0\) denotes the number of pixels with non-zero \(FV(x,y)\) values.

Local reconstruction errors were characterized using RSM. The underlying assumption is that, if the super-resolved reconstruction $I_H$ is faithful, an appropriately intensity-rescaled and blurred version of $I_H$ should reproduce the diffraction-limited wide-field image $I_L$ of the same field of view. Global intensity differences were compensated by applying a linear mapping $I_{HS}(\mu,\theta) = \mu I_H + \theta$, where $\mu$ and $\theta$ are a multiplicative gain and an additive offset that adjust overall brightness and background. The intensity-scaled image $I_{HS}$ was then convolved with an anisotropic Gaussian resolution-scaling function $I_{\mathrm{RSF}}(\sigma_x,\sigma_y)$ to mimic the optical blurring of the diffraction-limited system, yielding a pseudo wide-field image $I_{HL} = I_{HS}(\mu,\theta)\otimes I_{\mathrm{RSF}}(\sigma_x,\sigma_y)$, where $\otimes$ denotes convolution and $\sigma_x$, $\sigma_y$ are the Gaussian standard deviations along the $x$- and $y$-axes. The parameters $\mu$, $\theta$, $\sigma_x$ and $\sigma_y$ were jointly optimized by minimizing the squared $L_2$ norm between $I_{HL}$ and the measured diffraction-limited image $I_L$. After optimization, the resolution-scaled error map was defined pixel-wise as:
\begin{equation}
        \operatorname{RSM}(x,y)
        = \left| I_L(x,y) - I_{HL}(x,y) \right| 
\end{equation}
such that low RSM values indicate high local consistency between the super-resolved reconstruction and the diffraction-limited reference, corresponding to higher effective resolution and fewer reconstruction artefacts.

\subsection*{Funding}
This work was supported by National Natural Science Foundation of China (62405136, 62227818, 62405137, 62275125, 62275121, 12204239, 62175109), China Postdoctoral Science Foundation (BX20240486, 2024M754141), Youth Foundation of Jiangsu Province (BK20241466, BK20220946), Jiangsu Funding Programfor Excellent Postdoctoral Talent (2024ZB671), Fundamental Research Funds for the Central Universities (30922010313), Fundamental Research Funds for the Central Universities (2023102001), and Open Research Fund of Jiangsu Key Laboratory of Spectral Imaging $\&$ Intelligent Sense (JSGP202201, JSGPCXZNGZ202402).

\subsection*{Author contributions}
C.Z. and Q.C. supervised the project. C.Z. and J.Q. initiated and conceived the research. J.Q and J.F. programmed the reconstruction algorithm and analyzed the data. J.Q. and J.F. built the SIM system. H.W. helped with the biological experiment design. J.Q., J.F., M.Z., D.L., T.K., and X.H. performed the experiments. C.Z., J.Q., J.F. and H.W. wrote the manuscript with input from all authors.

\subsection*{Data and materials availability}
Source codes and data for 4I-SIM is publicly available at \textcolor{blue}{https://doi.org/xxxxxx}. Further details regarding the code can be found in Supplementary Note S7.

\subsection*{Conflict of interest}
The authors declare that they have no conflict of interest.

\clearpage

\begin{edfigure}
    \centering
    \includegraphics[width=1\linewidth]{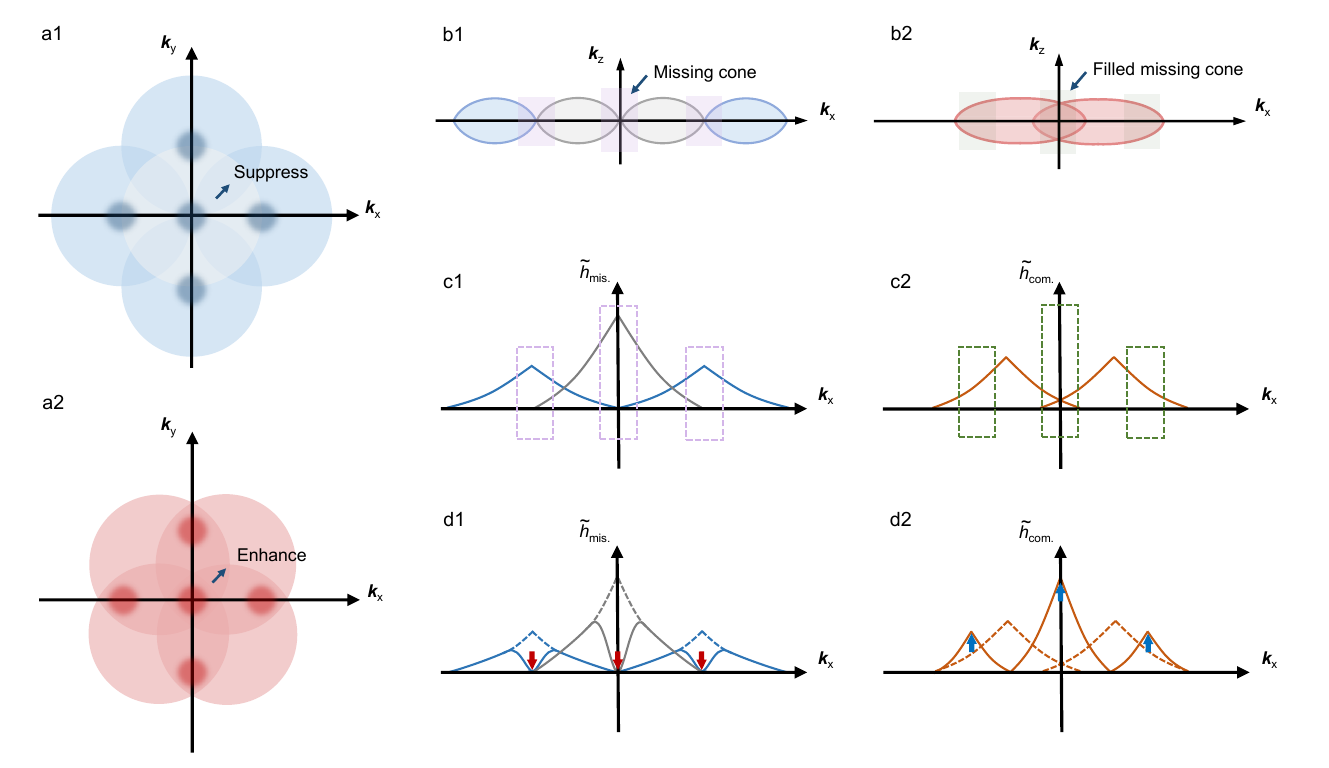}
    \caption{\textbf{Composite frequency-domain filtering constrained by the 3D OTF of 4I-SIM.} 
    \textbf{a1} Lateral support of the 0th and 2nd spectral orders in 4I-SIM, containing the missing-cone region. 
    \textbf{a2} Lateral support of the 1st spectral order in 4I-SIM, which compensates for the missing-cone region. 
    \textbf{b} Cross-sectional view of \textbf{a} in the \(\mathbf{k}_{xz}\) plane. From the 3D OTF perspective, \textbf{b1} exhibits a clear missing-cone effect, whereas \textbf{b2} covers the regions missing in \textbf{b1}. 
    \textbf{c} The 2D OTF obtained by projecting the 3D OTF in \textbf{b} along the \(\mathbf{k}_x\) axis. From the 2D OTF perspective, the spectral components projected from the missing-cone regions are emphasized, while information from regions with strong axial response is suppressed.
    \textbf{d} For effective 2D OS, information from the missing-cone regions should be suppressed while enhancing information that compensates for these missing components.}
    \label{EDF0}
\end{edfigure}

\begin{edfigure}
	\centering
	\includegraphics[width=1\linewidth]{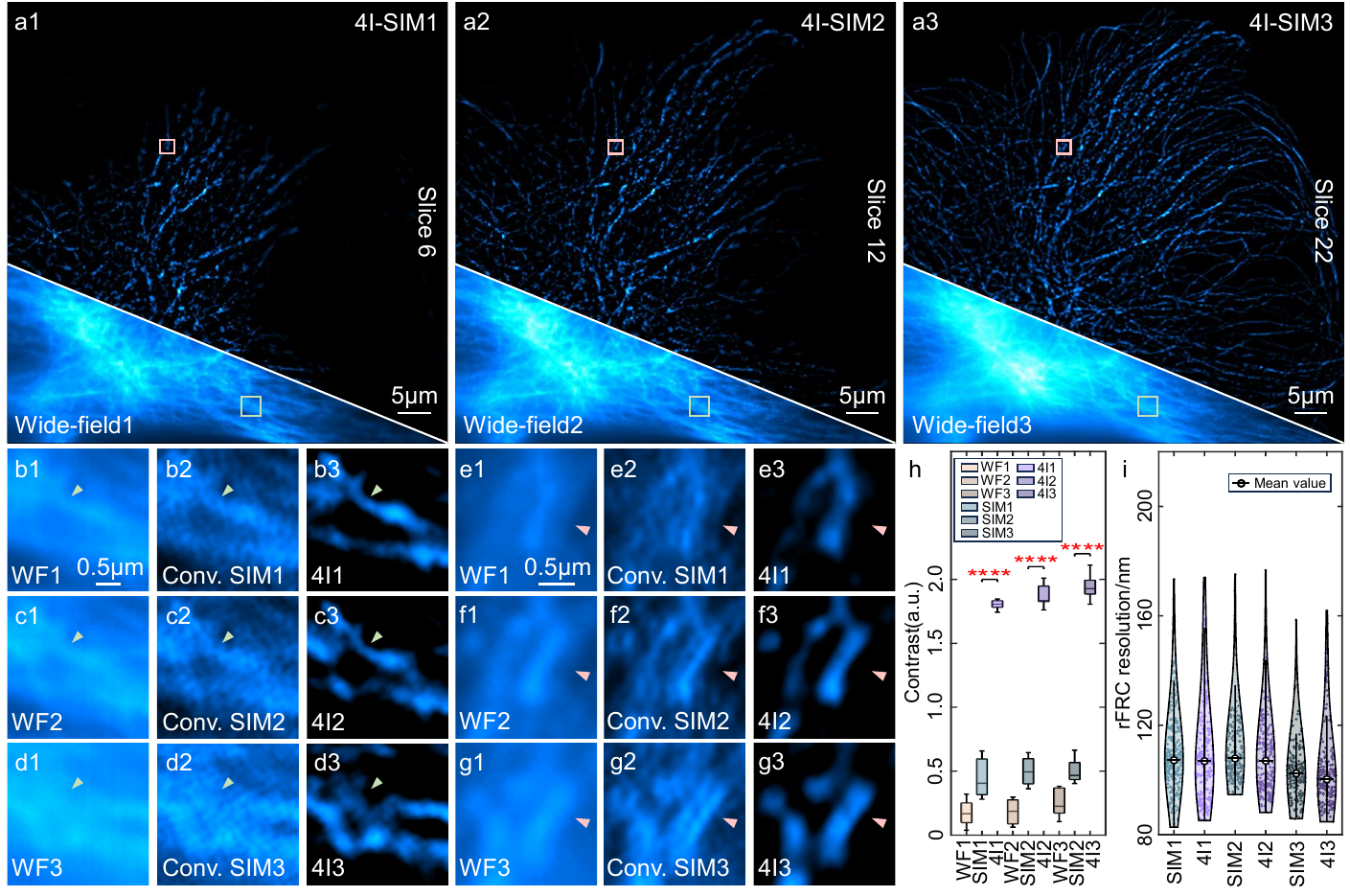}
		\caption{\textbf{Comparison of super-resolution imaging results of COS-7 cell microtubules labeled with BODIPY\textsuperscript{R} FL goat anti-mouse IgG.}  
		\textbf{a} Wide-field and 4I-SIM super-resolution images acquired at different axial planes. Raw SIM images were recorded at a resolution of $1024 \times 1024$ pixels using a 100$\times$ oil-immersion objective (UPlanXApo 100/1.45 Oil, Olympus, Japan).  
		\textbf{b}-\textbf{d} Magnified views obtained using different methods (wide-field, conventional SIM and 4I-SIM) from the green boxed region in \textbf{a}.
		\textbf{e}-\textbf{g} Magnified views obtained using different methods (wide-field, conventional SIM and 4I-SIM) from the pink boxed region in \textbf{a}.
		\textbf{h} Local contrast comparison across different axial planes and reconstruction methods.  
		\textbf{i} Lateral resolution distributions across axial layers, calculated using the rFRC method.  
		A two-tailed paired Student's $t$-test was applied to contrast values in \textbf{h}, with **** indicating $p < 0.0001$. Each experiment was independently repeated ten times with consistent results. Colored arrows indicate regions with pronounced reconstruction differences.  
		Scale bars: 5 $\mu$m (\textbf{a}); 500 nm (\textbf{b}-\textbf{g}).}
	\label{EDF1}
\end{edfigure}

\clearpage

\begin{edfigure}
    \centering
    \includegraphics[width=1\linewidth]{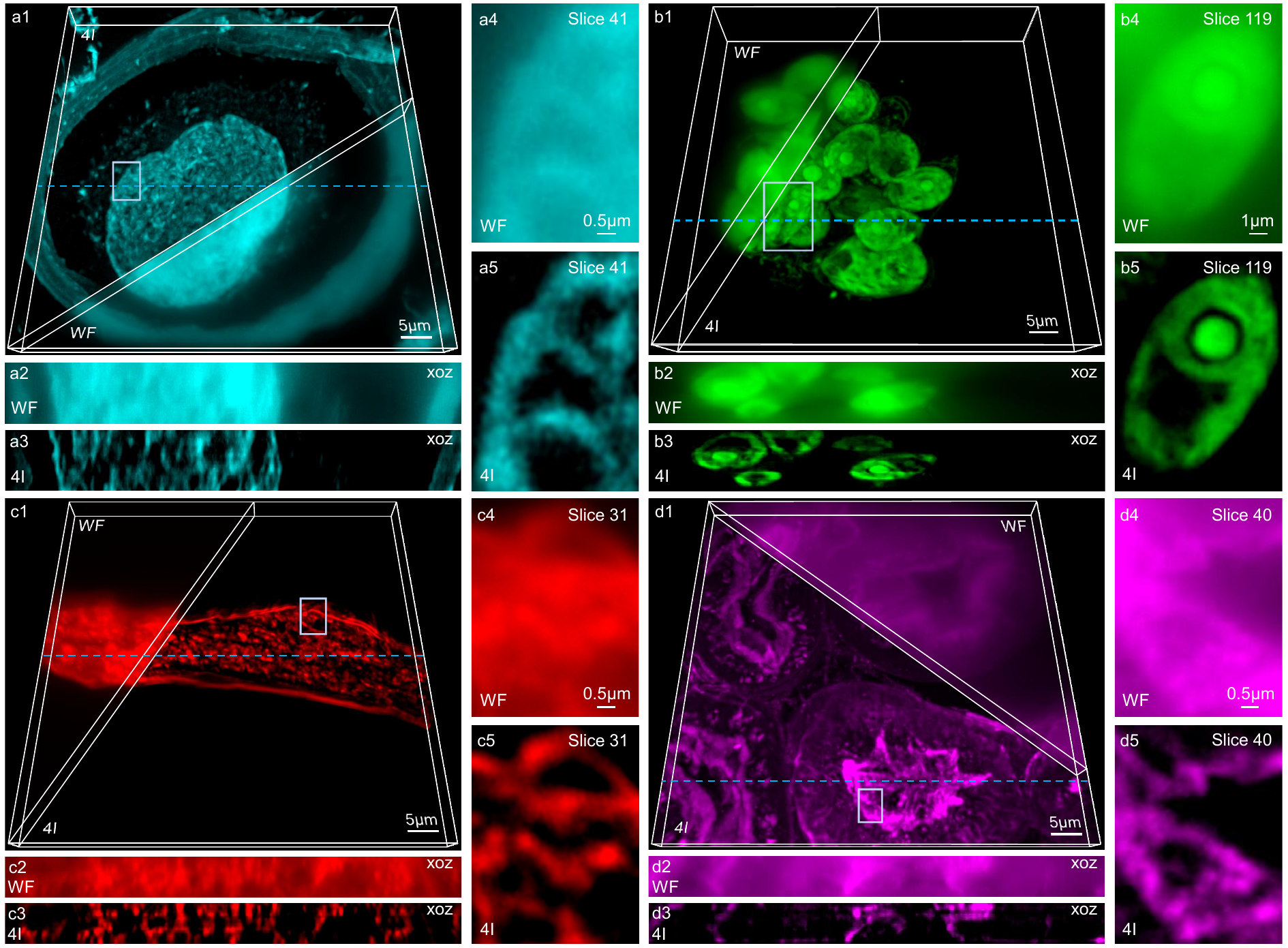}
    \caption{\textbf{Comparative experiments on different thick samples.} 
		\textbf{a1} Wide-field and 4I-SIM super-resolution images of the entire 3D field of view of a fixed autofluorescent \textit{Ascaris suum} fertilized egg.  
		\textbf{a2, a3} $x$-$z$ slices extracted from the wide-field and super-resolution images in \textbf{a1} along the blue dashed line.  
		\textbf{a4, a5} Enlarged view of the blue boxed region in \textbf{a1}, taken from slice 41, comparing wide-field and 4I-SIM.  
		\textbf{b} Wide-field and 4I-SIM super-resolution images of a fixed \textit{Chlamydomonas} sample.  
		\textbf{c} Wide-field and 4I-SIM super-resolution images of a fixed mouse brain section.  
		\textbf{d} Wide-field and 4I-SIM super-resolution images of a fixed mouse kidney section.  
		Raw SIM images were recorded at a resolution of $1024 \times 1024$ pixels using a 60$\times$ oil-immersion objective (UPlanXApo 100/1.45 Oil, Olympus, Japan).  
		Scale bars: 5 $\mu$m (\textbf{a1}, \textbf{b1}, \textbf{c1}, \textbf{d1}); 500 nm (\textbf{a4}, \textbf{a5}, \textbf{c4}, \textbf{c5}, \textbf{d4}, \textbf{d5}); 1 $\mu$m (\textbf{b4}, \textbf{b5}). Scale on the $z$-axis: 80 slices, 150 nm per slice (\textbf{a}); 160 slices, 150 nm per slice (\textbf{b}); 40 slices, 150 nm per slice (\textbf{c}); 60 slices, 150 nm per slice (\textbf{d}).}
    \label{EDF2}
\end{edfigure}

\clearpage
\bibliography{references}




\end{document}